\documentclass[useAMS,usenatbib]{mn2e}

\usepackage{natbib, aas_macros}

\usepackage{amsmath}
\usepackage[dvips]{graphicx}
\usepackage{color}
\usepackage{wrapfig}
\usepackage{amssymb}

\newcommand{\Msun}{M_{\odot}}
\newcommand{\Mh}{M_{\rm h}}

\newcommand{\rhoth}{\rho_{\rm th}}

\newcommand{\Mstar}{M_{\star}}
\newcommand{\sigmadla}{\sigma_{\rm DLA}}
\newcommand{\sigmalls}{\sigma_{\rm LLS}}
\newcommand{\dd}{{\rm d}}
\newcommand{\HI}{H\,{\sc i}}
\newcommand{\NHI}{N_{\rm HI}}
\newcommand{\SFR}{\rm SFR}
\newcommand{\Zsun}{Z_{\odot}}
\newcommand{\fn}{f(N_{\rm HI})}
\newcommand{\nuv}{n_{\rm th}^{\rm UV}}
\newcommand{\nh}{n_{\rm H}}
\newcommand{\Msunyr}{\rm M_{\odot}\,yr^{-1}}

\title[Effect of Radiative Transfer on DLAs and LLS]
{Effect of Radiative Transfer on Damped Lyman-$\alpha$ and Lyman Limit Systems in Cosmological SPH Simulations}
\author[Yajima et al.]
{Hidenobu Yajima$^{1, 2}$\thanks{E-mail:yuh19@psu.edu},  Jun-Hwan Choi$^{3}$,  Kentaro Nagamine$^{4}$\\
$^{1}$Department of Astronomy and Astrophysics, The Pennsylvania State University, 525 Davey Lab, University Park, PA 16802, USA\\
$^{2}$Institute for Gravitation and the Cosmos, The Pennsylvania State University, University Park, PA 16802, USA\\
$^{3}$Department of Physics and Astronomy, University of Kentucky, Lexington, KY, 40506-0055, U.S.A. \\
$^{4}$Department of Physics and Astronomy, University of Nevada, Las Vegas, 4505 S. Maryland Pkwy, Las Vegas, NV, 89154-4002, U.S.A.
}

\begin{document}


\pagerange{\pageref{firstpage}--\pageref{lastpage}} \pubyear{2008}

\maketitle

\label{firstpage}

%
%
\begin{abstract}

We study the effect of local stellar radiation and ultraviolet background radiation (UVB) on the physical properties of damped Lyman-$\alpha$ systems (DLAs) and Lyman limit systems (LLSs) at $z=3$ using cosmological SPH simulations.
We post-process our simulations with the Authentic Ray Tracing (ART) code for radiative transfer of local stellar radiation and UVB. 
We find that the DLA and LLS cross sections are significantly reduced by the UVB, whereas the local stellar radiation does not affect them very much except in the low-mass halos. 
This is because the clumpy high-density clouds near young star clusters effectively absorb most of the ionizing photons from young stars.
We also find that the UVB model with a simple density threshold for self-shielding effect can reproduce the observed column density distribution function of DLAs and LLSs very well, and we validate this model by direct radiative transfer calculations of stellar radiation and UVB with high angular resolution. 
We show that, with a self-shielding treatment, the DLAs have an extended distribution around star-forming regions typically on $\sim 10-30$ kpc scales, and LLSs are surrounding DLAs on $\sim 30-60$ kpc scales. 
The DLA gas is less extended than the virial radius of the halo, and LLSs are distributed over the similar scale to the virial radius of the host halo. 
Our simulations suggest that the median properties of DLA host haloes are:
$\Mh = 2.4 \times 10^{10} ~\Msun$, SFR$ = 0.3 ~ \Msunyr$, $M_{\star} = 2.4 \times 10^{8}~ \Msun$, and $Z / Z_{\rm \odot} =  0.1$. 
About 30 per cent of DLAs are hosted by haloes having $SFR = 1 - 20~\Msunyr$, which is the typical SFR range for LBGs.
More than half of DLAs are hosted by the LBGs that are fainter than the current observational limit.
Our results suggest that fractional contribution to LLSs from lower mass haloes is greater than for DLAs.
Therefore the median values of LLS host haloes are somewhat lower with 
$\Mh = 9.6 \times 10^{9} ~\Msun$, SFR$ = 0.06 ~ \Msunyr$, $M_{\star} = 6.5 \times 10^{7}~ \Msun$ and $Z / Z_{\rm \odot} =  0.08$. 
About 80 per cent of total LLS cross section are hosted by haloes with ${\rm SFR} \lesssim 1~\Msunyr$, hence 
most LLSs are associated with low-mass halos with faint LBGs below the current detection limit. 

\end{abstract}

%
%

%
%
\section{INTRODUCTION}
\label{sec:intro}
Interplay between neutral hydrogen gas and star formation in galaxies is a crucial issue for understanding galaxy 
formation and evolution. 
Absorption systems in the spectrum of QSOs are good tracers of neutral hydrogen, 
and large surveys such as the Sloan Digital Sky Survey (SDSS) provide abundant data on \HI\  gas at high redshift 
\citep{Prochaska05-1, Prochaska09}. 
The absorption systems with $\NHI \geq 2\times 10^{20}\,{\rm cm}^{-2}$ show the damping wing in the absorption 
profile, and are called Damped Ly$\alpha$ systems (DLAs) \citep{Wolfe86}.
In addition, recently the DLAs has been studied even by GRB aftergrow \citep[e.g.,][]{Prochaska07b, Ledoux09, 
Rau10}.  
It is known that DLAs dominate the neutral gas content of Universe between $z \sim 0 - 5$
 \citep{Wolfe86, Storrie96, Storrie00}.  
 The global neutral gas content measured from DLAs declines with cosmic epoch in concert with the growth in 
cosmic stellar mass \citep{Storrie00}. 
 Therefore, it is generally thought that DLAs are strongly correlated with cosmic star formation.
However, in spite of numerous observational sample of DLAs, a deep understanding of DLAs and host galaxies 
remains elusive.

 The nature of DLAs has been revealed gradually by recent progress in observations 
 \citep[e.g.,][]{Wolfe95, Storrie00, Rao00, Peroux03, Chen03, Prochaska05, Prochaska09, Prochaska10a, Wolfe08, 
Noterdaeme09}
 and simulations
 \citep[e.g.,][]{Katz96, Gardner97a, Gardner97b, Haehnelt98, Maller01, Gardner01, Razoumov06a, Nagamine04f-1, 
Nagamine04g-1, Nagamine07, Pontzen08,  Barnes09, Tescari09, Cen10}.
Thanks to these works, our understanding on the relationship between DLAs and host galaxies have improved. 
For example, it has been suggested that a significant fraction of DLAs are distributed in the same dark matter halos 
as the Lyman break galaxies \citep[LBGs; e.g.,][]{Adelberger05, Cooke06, Rafelski11}.
Theoretically, \citet{Cen10} examined the number fraction of DLAs related to LBGs using cosmological AMR 
simulations.
\citet{Lee11} estimated the DLA--LBG cross-correlation function using cosmological SPH simulations, and showed 
that their result agrees well with the observational estimate of \citet{Cooke06}. 

A connection between DLAs and Ly$\alpha$ emitters (LAEs) has also been suggested. 
\citet{Rahmani10} investigated Ly$\alpha$ emission from 341 DLAs at $z\sim 2.86$, and concluded that the overall 
DLA population seems to originate from the low luminosity end of LAEs at high redshift.   
\citet{Barnes11} simulated three dimensional Ly$\alpha$ radiative transfer (RT) for high-redshift galaxies in 
cosmological SPH simulations. They showed that DLA host galaxies show extended Ly$\alpha$ emission around 
high-column density region.
The Ly$\alpha$ emission is extended over several arc seconds, and has the spectral width of several hundred km
\,s$^{-1}$.
A combination of these observational and theoretical works seem to suggest that DLAs are associated with a wide 
variety of sources, such as LBGs and LAEs. 

For a better understanding of DLAs, it is quite important to investigate various feedback effects in galaxies which 
can drastically change the state of gas. 
Multiple supernovae can cause galactic wind, and hence change gas distribution largely in galaxies. 
\citet{Nagamine04f} investigated the effect of galactic wind on the \HI\ column density distribution function $\fn$, 
and found that strong galactic winds with a speed of $\sim$500\,km\,s$^{-1}$ is needed to reproduce the observed 
$\fn$ at high $\NHI$ values.  
\citet{Tescari09} also found that strong galactic winds with speed of $\sim$600\,km\,s$^{-1}$
 are needed in order to reproduce the observed $\fn$ and the evolution of \HI\ mass density with redshift.  
Several authors have suggested that galactic outflows can produce the large velocity dispersion as observed in 
DLAs \citep[e.g.,][]{Razoumov09, Hong10, Cen10}. 
 
 Gas inflows from surrounding IGM can also have strong impact on the absorption systems in high-$z$ galaxies.  
Recent simulations have revealed a new theoretical paradigm for galaxy evolution that
 a large amount of gas penetrate deep inside dark matter haloes as cold, filamentary streams
 \citep{Katz03, Keres05, Keres09, Birnboim03, Dekel06, Ocvirk08, Brooks09, Dekel09}.
 \citet{Faucher11} studied the neutral hydrogen content in the cold accretion 
 to LBGs using SPH simulations with UVB RT. 
 They showed that a part of gas could be DLAs, however the covering fraction was quite small.
 Similarly, \citet{Fumagalli11} showed the neutral hydrogen gas content in cold stream by 
 cosmological AMR simulations and RT calculation of local stellar source and UVB.
 They find that most of the gas in cold streams is ionized, however it can survive as Lyman-limit systems (LLSs), 
which is defined by column densities of $10^{17} < \NHI < 10^{20.3}$ atoms cm$^{-2}$.
  
  The observations of DLA metallicity can also give important constraints on feedback processes, as the interstellar 
gas can be enriched by supernovae in the early Universe.
  \citet{Moller04} presented the metallicity distribution of DLAs and suggested that the DLAs with Ly$\alpha$ 
emission tend to have higher metallicities of $Z \gtrsim 10^{-1.5}~\Zsun$.
 \citet{Ledoux06} presented the velocity width $\Delta v$--metallicity relation of DLAs from the line width of  low-
ionization metals, and showed that the metallicity is linearly proportional to $\Delta v$ with a large dispersion.
 They also suggested that the slope of fitting function for $\Delta v$--metallicity relation is the same for the sample 
at  $z > 2.43$ and $z \le 2.43$, but the median metallicity decreases with increasing redshift.
 Moreover, \citet{Prochaska07a} presented the metallicities of a larger DLA sample with a large dispersion similar to 
that of \citet{Ledoux06}.
  Theoretically, recent numerical simulations which include metal enrichment by supernovae have successfully 
produced the observed metal distribution of DLAs \citep{Pontzen08, Tescari09, Cen10}.   
 
 Another important feedback on DLAs is radiation from UVB and local stellar sources.
 The UV radiation can ionize hydrogen gas, and largely reduce the neutral fraction of ISM. 
Earlier \citet{Katz96} and \citet{Haehnelt98} included the self-shielding correction for UVB radiation 
to reproduce DLA statistics in there numerical simulations.  Thereafter,
recent simulations have reproduced various observation data of DLAs reasonably well
by considering the effect of UV RT 
\citep{Razoumov06a, Razoumov08, Kohler07, Pontzen08, Nagamine10-1, Altay11, Faucher11, Fumagalli11, McQuinn11}.
 Most of them have focused on the RT of UVB, and found that the self-shielding effect is
 crucial for matching the observed $\fn$ of DLAs.
\citet{Nagamine10} showed that the simple self-shielding model without RT calculations 
could reproduce the observed $\fn$ very well except for the very high $\NHI$ values. 
 \citet{Altay11} also reproduced the observed $\fn$ in the range of $10^{12} < \NHI  < 10^{22}$\,atoms\;cm$^{-2}$ 
by post-processing cosmological SPH simulations for the UVB RT. 

On the other hand, the flux of stellar radiation near the star cluster can be much higher than the local UVB,
hence stellar RT can be important for DLAs and LLSs.
However, it has been ignored in many simulations.
Some authors have suggested that the UV radiation from local stellar sources can be important for absorption systems \citep{Miralda05, Schaye06}, however, there have not been extensive studies of direct RT calculations that examined the effect of local stellar sources. 
\citet{Nagamine10} showed that the $\fn$ at $z=3$ does not change very much, based on a direct RT calculation 
from young stellar sources. 
\citet{Fumagalli11} carried out the RT calculations of local stellar sources and UVB for seven high-resolution simulated galaxies, and found that the local stellar radiation changed $\fn$ at $\NHI \sim 10^{18}-10^{21}$\;cm$^{-2}$ by $\sim$0.5 dex to the other model without local stellar radiation.
Their simulation result is in good agreement with observation at $\log \NHI > 21$, however it underestimates $\fn$ 
at lower $\NHI$ range of LLS. 

The stellar radiation flux onto a gas cloud varies depending on the distance between gas and stars and the type of SED.
Therefore in a more realistic inhomogeneous medium, 
precise RT calculations are needed to study the effects of stellar radiation on absorption systems.
Furthermore, since the distribution of stars and gas varies among different galaxies \citep[e.g.,][]{Yajima11},
we need to carry out the RT calculation for many galaxies to discuss the statistical data such as $\fn$.
In the present work, we perform precise RT calculations with $\sim 10^{4}$ angular bins for each star particle in cosmological SPH simulations, and examine the distribution of neutral hydrogen gas for a few hundred galaxies.  
Our RT code is based on the ray-tracing method, which can estimate the ionization structure more accurately than the Monte Carlo method that \citet{Fumagalli11} used \citep[see][for a comparison of RT methods]{Iliev06}.
In addition, the number of galaxy sample processed with RT in our work is much larger ($N \gtrsim 100$) than that of \citet[][$N\sim 8$]{Fumagalli11}. 
Thus, our calculation allows us to investigate the effect of RT more reliably on the statistical quantities of DLAs and LLSs, such as $\fn$ and the mass dependence of cross sections.
Moreover, by direct RT calculations of UVB, we verify the self-shielding model of UVB derived by \citet{Nagamine10}.   
We also discuss the physical quantities of host galaxies of DLAs and LLSs, such as the halo mass, star formation 
rate, stellar mass and metallicity.

Our paper is organized as follows.
We describe our simulations and our approach for RT calculations in Section~\ref{sec:model}. 
In Section~\ref{sec:result}, we present our results, and show the effect of stellar and UVB radiation on cross sections of DLAs and LLSs, and on $\fn$. 
In Section~\ref{sec:abundance}, the cumulative abundance of DLAs is shown.
In addition, we show the typical physical quantities of DLAs and LLSs host galaxies.
Finally, we summarize our main conclusions in Section~\ref{sec:summary}.
We focus on redshift $z=3$ in this paper.

%
%
\section[]{MODEL AND METHOD}
\label{sec:model}

\subsection{Simulations}

We use an updated and modified version of the Tree-particle-mesh (TreePM) smoothed particle hydrodynamics (SPH) code GADGET-3 \citep[originally described in][]{Springel05e}.  The SPH calculation is performed based on the entropy conservative formulation \citep{Springel02}.  Our fiducial code includes radiative cooling by H, He, and metals \citep{Choi09a}, star formation, supernova feedback, a phenomenological model for galactic winds and a sub-resolution model of multiphase ISM 
\citep{Springel03a}.  We also include the heating by a uniform UVB, which we will discuss more in Section~\ref{sec:UVB}.

In this multiphase ISM model, high-density ISM is pictured to be a two-phase fluid consisting of cold clouds in pressure equilibrium with a hot ambient phase.  Cold clouds grow by radiative cooling out of the hot medium, and this material forms the reservoir of baryons available for star formation. The star formation rate (SFR) is estimated for each gas particle that have densities above the threshold density, and the star particles are spawned statistically based on the SFR. For the star formation model, the ``Pressure SF model'' is being used \citep{Schaye08, Choi09b}. This model estimates the SFR based on the local gas pressure rather than the gas density, and prevents artificial fragmentation of the gas. 

The simulations used in this paper also uses the ``Multicomponent Variable Wind" model developed by \citet{Choi10}. This wind model is based on both the energy-driven wind and the momentum-driven wind \citet{Murray05},  and the wind speed depends on the galaxy stellar mass and SFR.  It gives more favorable results when compared to the observations of cosmic C\,{\sc iv} mass density and IGM temperature than the previous model with a constant wind speed.  To enable this new wind model, \citet{Choi10} implemented an on-the-fly group-finder into GADGET-3 to compute galaxy masses and SFRs while the simulation is running.  The group-finder, which is a simplified variant of the \textup{SUBFIND} algorithm developed by  \citet{Springel01}, 
identifies the isolated groups of star and gas particles (i.e., galaxies) based on the baryonic density field. The detailed procedure of this galaxy grouping is described in \citet{Nagamine04e}. 

The simulations used in this paper are performed with $2\times144^{3}$ particles. 
The DM and gas particles masses are $1.97\times10^{7}$ and $4.06\times10^{6}\;h^{-1} \Msun$, respectively. 
The comoving gravitational softening length is $\epsilon = 2.78 h^{-1}$\,kpc in a comoving $(10 h^{-1}\rm Mpc)^{3}$ calculation box.

The adopted cosmological parameters are consistent with the WMAP results \citep{Komatsu11}:
$H_0 = 72$\,km\,s$^{-1}$\,Mpc$^{-1}$ ($h=0.72$), $\Omega_M=0.26$, $\Omega_{\Lambda}=0.74$, $\Omega_b=0.044$, $\sigma_8 = 0.80$, and $n_s = 0.96$.

\subsection{Radiation Transfer}

\subsubsection{Stellar Radiation}
\label{sec:radiation}

To study the effect of stellar radiation on DLAs and LLSs, we compute the stellar radiation transfer 
and the ionization structure of gas in each dark matter halo by post-processing the simulation output.
The RT scheme used in this paper is the Authentic Radiation Transfer (ART) method, and the treatment is basically 
the same as in \citet{Yajima09, Yajima11-1, Yajima11b}.
Here we briefly summarize the procedure of RT calculation. 

First we set up a uniform grid around each dark matter halo with a grid cell size equal to the gravitational softening 
length of the simulation, and translate the SPH gas information into a gridded data.
 As for the scattering of photons, we employ the on-the-spot approximation \citep{Osterbrock89}, in which the 
scattered photons are assumed to be absorbed immediately on the spot.
In this work, the RT equation is solved along $n_{\rm g}^{2}$ rays ($\sim$10$^{4-5}$) with a uniform angular resolution from each source, where $n_{\rm g}$ is the grid number on a side of the calculation box of each galaxy.
We shoot the radiation rays in a radial fashion from each star particles. 
The number of ionizing photons emitted from the source stars is computed based on the theoretical spectral energy 
distribution (SED) given by the population synthesis code P\'{E}GASE v2.0 \citep{Fioc97}.
We consider the effect of age and metallicity of the stellar population by interpolating the table generated from the 
result of P\'{E}GASE.  
We assume the \citet{Salpeter55} initial mass function with the mass range of $0.1 - 50 \Msun$.

In addition, we consider the dust extinction.
The dust model is similar to \citet{Yajima11}, e.g.,  
$m_{\rm d} = 0.01 \times m_{\rm g} \frac{Z}{Z_{\odot}}$ with a silicate-like mass density of 3\,g\,cm$^{-3}$, where 
$m_{\rm d}$, $m_{\rm g}$ and $Z$ are dust mass, gas mass and metallicity.
The size distribution is $n_{\rm d}(a_{\rm d}) \propto a^{-3.5}_{\rm d}$ \citep{Mathis77}
with the dust radius of $0.1 - 1.0~\rm \micron$.
Since the assumed range of dust size is larger than the wavelength of Lyman limit, we assume 
the absorption efficiency $Q(\nu)=1$ for the ionizing photons \citep{DL84}.

\subsubsection{UV Background Radiation}
\label{sec:UVB}

The UVB can ionize the baryonic gas in galaxies, and then gas can be heated up to $\sim 10^4$\,K.
It would be ideal to compute the RT of both UVB and stellar radiation as the simulation runs, however in practice it is a very expensive calculation.  In general, the calculation amount of UVB transfer is larger than that of stellar radiation by $2-4$ orders of magnitude \citep{Yajima11}.  
%
Therefore we first use the simple, four different models of UVB, similarly to \citet{Nagamine10}.
We then compute the actual UVB transfer for some galaxies to check the validity of the UVB models. 

Our fiducial simulations include a uniform UVB with a modified \citet{Haardt96} (hereafter HM96) spectrum \citep[see][]{Dave99}, where the reionization takes place at $z\simeq 6$ as suggested by the quasar observations \citep[e.g.,][]{Becker01} and stellar radiative transfer calculations \citep[e.g.,][]{Sokasian03}.

We use the following four simulations with different treatment of UVB (but with the same initial condition) to examine the effects of UVB:
\begin{enumerate}
\item {\bf Optically-Thin Model}:
A uniform UVB radiation field with an optically thin approximation. \\
\item {\bf HM0.5 Model} (HM96 $\times\; 0.5$): 
The ISM is optically thin to the same UVB, however the intensity of UVB is reduced to the half of the Optically-Thin model. \\
\item {\bf OTUV Model} (Optically Thick UV):
The ISM is optically thin to the UVB in the lower density regions with $n_{\rm H} < 0.01 \rhoth = 6.34 \times 10^{-3}$\,cm$^{-3}$, but completely optically thick in higher density regions of $n_{\rm H} \ge 0.01 \rhoth$, where $ \rhoth$ is the threshold density, above which star formation is allowed. The value of $\rhoth$ was determined by \cite{Choi09b} based on the observed SF cut-off column density of the 
Kennicutt law.  The OTUV model implicitly assumes that the UVB cannot penetrate into the high density regions by self-shielding. 
This type of self-shielding model was first proposed by \citet{Katz96} and \citet{Haehnelt98} to examine $\fn$ and DLA kinematics.  
Later \citet{Nagamine10} successfully reproduced the observed $\fn$ by incorporating the OTUV model into cosmological simulations.   See also \citet{Pontzen08, Fumagalli11} and \citet{Altay11}. 
 \\
\item {\bf No-UVB Model}: UVB does not exist at all.
\end{enumerate}

Since we use the same initial condition for these runs, we can compare the effect of radiation halo by halo basis. 
After discussing the above four models in the beginning of Section~\ref{sec:fn}, we then move on to discuss the effects of radiative transfer in Sections~ \ref{sec:UVBtransfer} and \ref{sec:stellarRT}.  We compute the transfer of local stellar radiation in each halo as post-processing, and estimate the ionization structure in the equilibrium state by evaluating 
\begin{equation}
(\Gamma_{\rm UVB}^{\gamma}+ \Gamma_{\rm star}^{\gamma}) \; n_{\rm H_I}
+ \Gamma^{\rm C}\; n_{\rm H_I} \; n_{\rm e} = \alpha_{\rm B}\; n_{\rm H_{II}} \; n_{\rm e},
\end{equation}
where $\Gamma_{\rm UVB}^{\gamma}$ and $\Gamma_{\rm star}^{\gamma}$
are the photoionization rate by UVB and stellar radiation respectively;
$\Gamma^{\rm C}$ is the collisional ionization rate;
$n_{\rm e}, n_{\rm HI}$ and $n_{\rm HII}$ are the number density of
free-electron, neutral and ionized hydrogen, respectively;
$\alpha_{\rm B}$ is the total recombination coefficient to all bound excitation levels.
The value of $\Gamma_{\rm star}^{\gamma}$ is estimated by the full RT calculation,
but $\Gamma_{\rm UVB}^{\gamma}$ is computed with the above UVB model.


%
%
\section[]{RESULTS}
\label{sec:result}

We first compare our simulation results to the recent observations of \HI\ column density function \citep[e.g.,][]{Prochaska09, Noterdaeme09, Prochaska10b}, and examine the effect of UVB and stellar radiation on $\fn$. 
We then examine the \HI\ cross section of DLAs and LLSs, and study how they change with the UVB models and the RT calculation.  Finally, we study the typical physical properties of DLA and LLS gas, such as SFR and metallicity.

\subsection{Column density distribution function}
\label{sec:fn}

The column density distribution function $f(\NHI, X(z))$ is defined such that $f(\NHI, X) \dd \NHI \dd X$ is the number of absorbers per sight line with \HI\ column density in the interval $[\NHI, \NHI + \dd \NHI]$, and absorption distance
in the interval $[X, X + \dd X]$.
The absorption distance $X(z)$ is given by
\begin{equation}
X(z) = \int_{\rm 0}^{z} (1 + z')^{2} \frac{H_{\rm 0}}{H(z')} ~\dd z'.
\end{equation}
In practice, if the comoving box-size of the simulation is $\Delta L$, then the corresponding absorption distance per sight-line is 
\begin{equation}
\Delta X = \left( \frac{H_{\rm 0}}{c} \right) (1 + z)^{2} \Delta L.
\end{equation}
For example, for $\Delta L = 10 ~h^{-1} \rm~Mpc$ and $z=3$, we have $\Delta X = 0.0534$.

Since the simulation box size is very small, we can safely assume that the DLA clouds do not overlap along a sight-
line, and we can compute the $\NHI$ distribution function by counting the number of projected grid-cells with column densities in the range $[\NHI, \NHI + \dd \NHI]$ for all halos in the simulation.  We use the same grid that we set around each dark matter halo for the RT calculation, with the grid cell size equal to the comoving gravitational softening length.

\begin{figure}
\begin{center}
\includegraphics[scale=0.45]{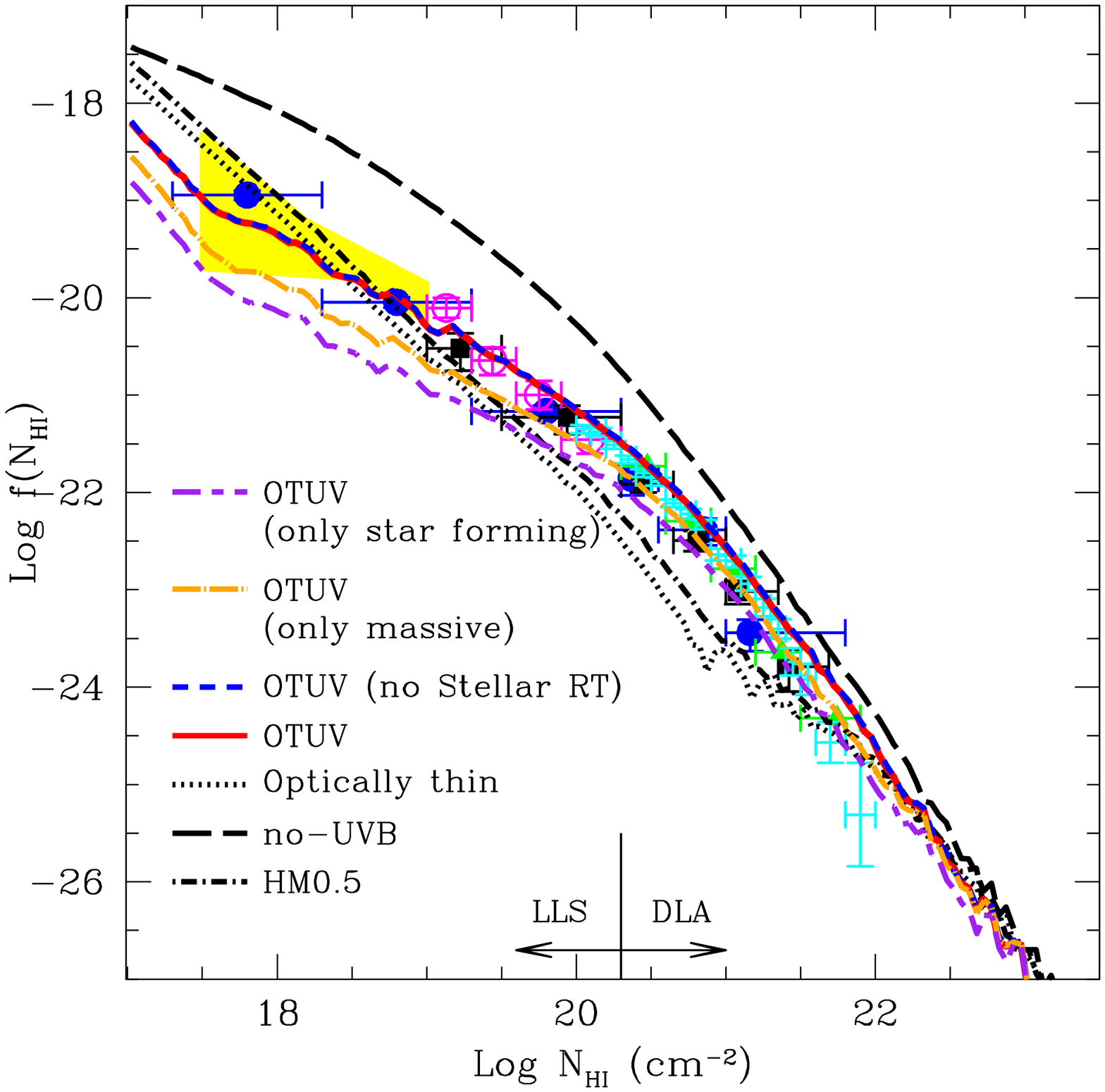}
\caption{
Column density distribution function at $z=3$.
The purple dash-long dashed line indicates the $\fn$ considering only star-forming galaxies that have young stellar sources ($t_{\rm age} \le 10^{7}~\rm yr$).
The orange dot-dash line indicates the $\fn$ considering only massive galaxies of $M_h \ge 5 \times 10^{10}~\Msun$.
The observation data points are from \citet[][blue filled circles,]{Peroux01},
 \citet[][black filled squares]{Peroux05},
\citet[][magenta open circles]{OMeara07}, 
\citet[][green triangles]{Prochaska09},
\citet[][cyan bars]{Noterdaeme09}
and \citet[][yellow shade]{Prochaska10b}.
The figure is similar to Figure 3 in \citet{Nagamine10}, but the $\NHI$ range is extended to a lower value of $\NHI  
= 10^{17}$\,cm$^{-2}$ to discuss LLS.
}
\label{fig:cdfall}
\end{center}
\end{figure}

We calculate the \HI\ density in  each grid-cell of volume $\left(\epsilon/[1+z]\right)^{\rm 3}$ after the RT calculation, 
and estimate $\NHI$ by projecting the \HI\ mass distribution along the $z$-axis direction as
\begin{equation}
N_{\rm HI} = \sum_{i} \frac{\epsilon\, \rho_{{\rm HI}, i}} {m_{\rm p} (1+z)}
\end{equation}
where $\epsilon$ is the comoving gravitational softening length, $m_{\rm p}$ is the proton mass, and $z$ is the redshift.  The resulting $\fn$ is insensitive to the direction of projection. 
By this method, we are treating each projected grid-cell element as one line-of-sight.
We have also checked that we are not missing any LLS columns in the outskirts of the halo by doubling the size of the grid for some of the halos.

Figure~\ref{fig:cdfall} shows $\fn$ in the range of $17 < \log \NHI < 22$ at $z=3$ together with some observational data. 
The no-UVB run overpredicts the observed data points at all $\NHI$ range. 
In particular, the difference becomes as large as an order of magnitude at $\NHI \sim 10^{19}$\,cm$^{-2}$. 
This shows that the ionization effect by the UVB is crucial in reproducing the observed LLS and DLA column density 
distribution functions. 
On the other hand, the $\fn$ of the Optically-Thin and the HM0.5 runs are close to the observed data at $\NHI \lesssim 10^{19.5}~\rm cm^{-2}$, however, falls below the observation data at $10^{19.5} ~{\rm cm^{-2}} \lesssim \NHI \lesssim 10^{21} ~{\rm cm^{-2}}$ by a significant amount. 

The OTUV run agrees with the observed $\fn$ much better than the other models in a wide range of $10^{17} ~{\rm cm^{-2}} \lesssim \NHI \lesssim 10^{\rm 21}~{\rm cm^{-2}}$.
At lower column densities of $\NHI \lesssim 10^{18}$\,cm$^{-2}$, the OTUV run is lower than the Opticall-Thin run, but it still fits with the observed range indicated by the yellow box. 
Thus, as far as the column density range of LLS and DLA is concerned, we conclude that the UVB treatment of the OTUV run successfully reproduces the observational data in a wide range of $10^{17} ~{\rm cm^{-2}} \lesssim \NHI \lesssim 10^{\rm 21}~{\rm cm^{-2}}$.
We note that the result of the OTUV run in Figure~\ref{fig:cdfall} is slightly different from that in \citet{Nagamine10}, but this is because the galactic wind model was slightly different between the two works. 
We have also checked the results with the updated UVB model of \citet{Faucher09}, and found that the results are not very different.

One notable feature in Figure~\ref{fig:cdfall} is that the OTUV run is higher than the observed $\fn$ at $\NHI \gtrsim 10^{21.5}$.
These high column density sight-lines mainly come from haloes of higher dust-to-gas ratio. 
In such halos, the molecular hydrogen H$_{2}$ is created efficiently, and the H$_2$ mass fraction can be $\gtrsim 0.1$ \citep[e.g.,][]{Hirashita05}. 
Because our current simulations do not include H$_{2}$ formation, we may overpredict $\fn$ at high column densities of $\NHI \gtrsim 10^{21.5}$\,cm$^{-2}$.

\begin{figure}
\begin{center}
\includegraphics[scale=0.43]{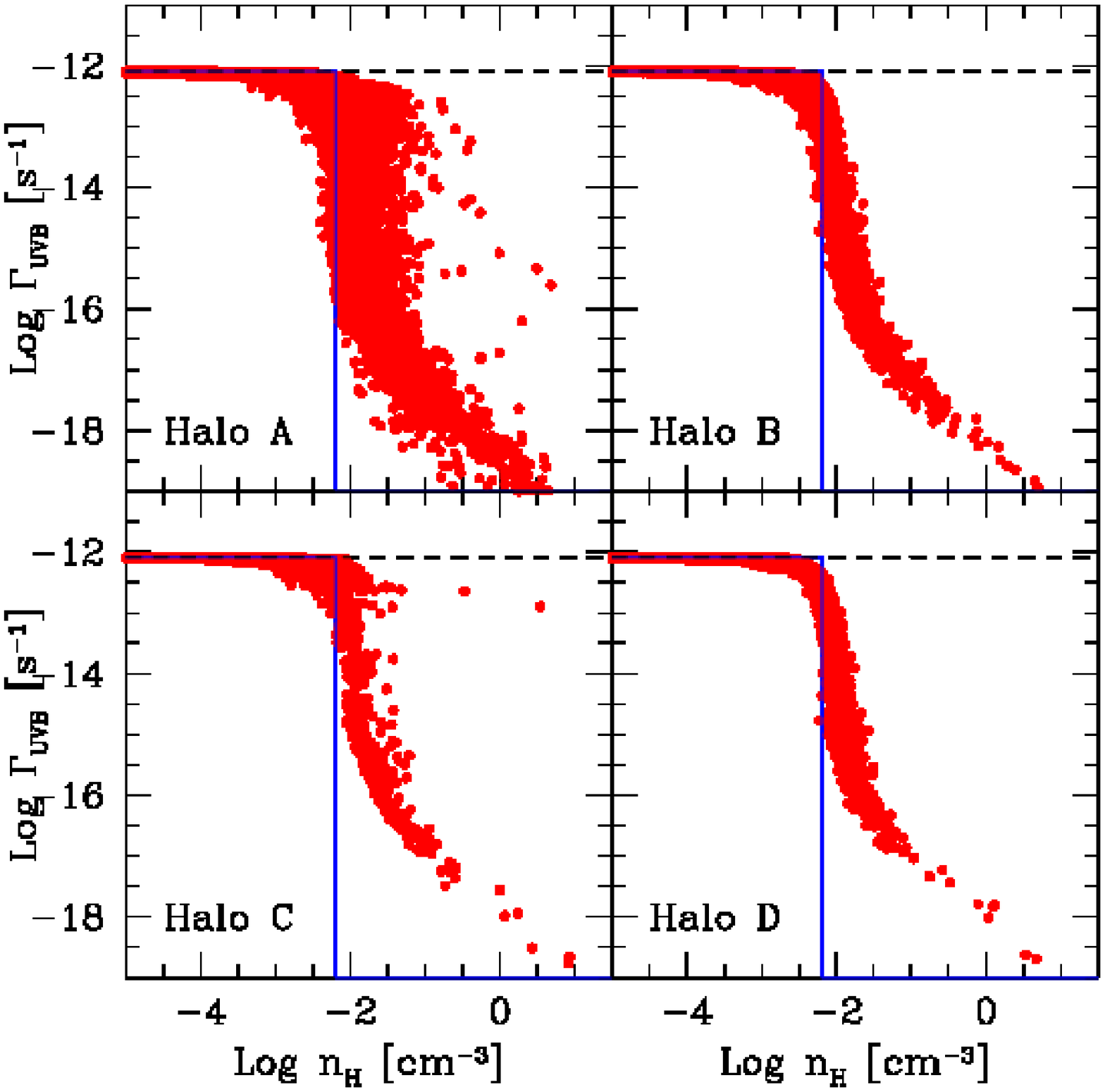}
\caption{
Photoionization rate by UVB radiation ($\Gamma_{\rm UVB}$) as a function of hydrogen number density for four different halos in the OTUV run (Halo A : $\Mh = 9.7\times 10^{10} \Msun$, B : $\Mh = 2.5 \times 10^{10}\Msun$, C : $\Mh = 1.5 \times 10^{10} \Msun$, and D : $\Mh = 9.4 \times 10^{9}\Msun$).
The halos are taken from the OTUV run, but we are solving the transfer of UVB radiation directly in each cell as shown by the red points.  The black dashed lines and the blue solid lines indicate the ionization rate in the Optically-Thin model and the OTUV model, respectively.
}
\label{fig:gammauvb}
\end{center}
\end{figure}


We find that the stellar radiation transfer does not affect the shape of $\fn$ (overlapping blue dashed line and red solid line in Figure~\ref{fig:cdfall}),
 whereas the UVB model changes the $\fn$ significantly. 
We will discuss the effect of stellar radiation on DLAs and LLSs in more detail later in the paper.

The change of column density by UVB in the OTUV run is similar to those found by previous works \citep[e.g., Fig.~A1 in][]{Fumagalli11}.  On the other hand, the effect by stellar radiation in our simulation is much smaller than that of \citet{Fumagalli11}.  
There could be several possible reasons for this difference as follows. 
In our simulations, the high-density gas clumps near young stars can effectively absorb stellar radiation,  and block the propagation of stellar radiation over extended regions in a galaxy \citep{Yajima11}.
Another possible reason could be different numerical methods: our RT method is based on a ray-tracing method, whereas Fumagalli's method is Monte Carlo.  The ray-tracing method can estimate the ionisation structure by point sources more accurately, while the Monte Carlo method can cause some artificial inhomogeneity \citep[see the comparison work by][]{Iliev06}.  
Finally, the spatial resolution is always a concern, and we note that the resolution of Fumagalli's AMR simulation in high-density regions is better than ours.  Unfortunately it is very difficult to tell the exact resolution effects unless we compare the two codes closely side by side. 

\begin{figure*}
\begin{center}
\includegraphics[scale=0.4]{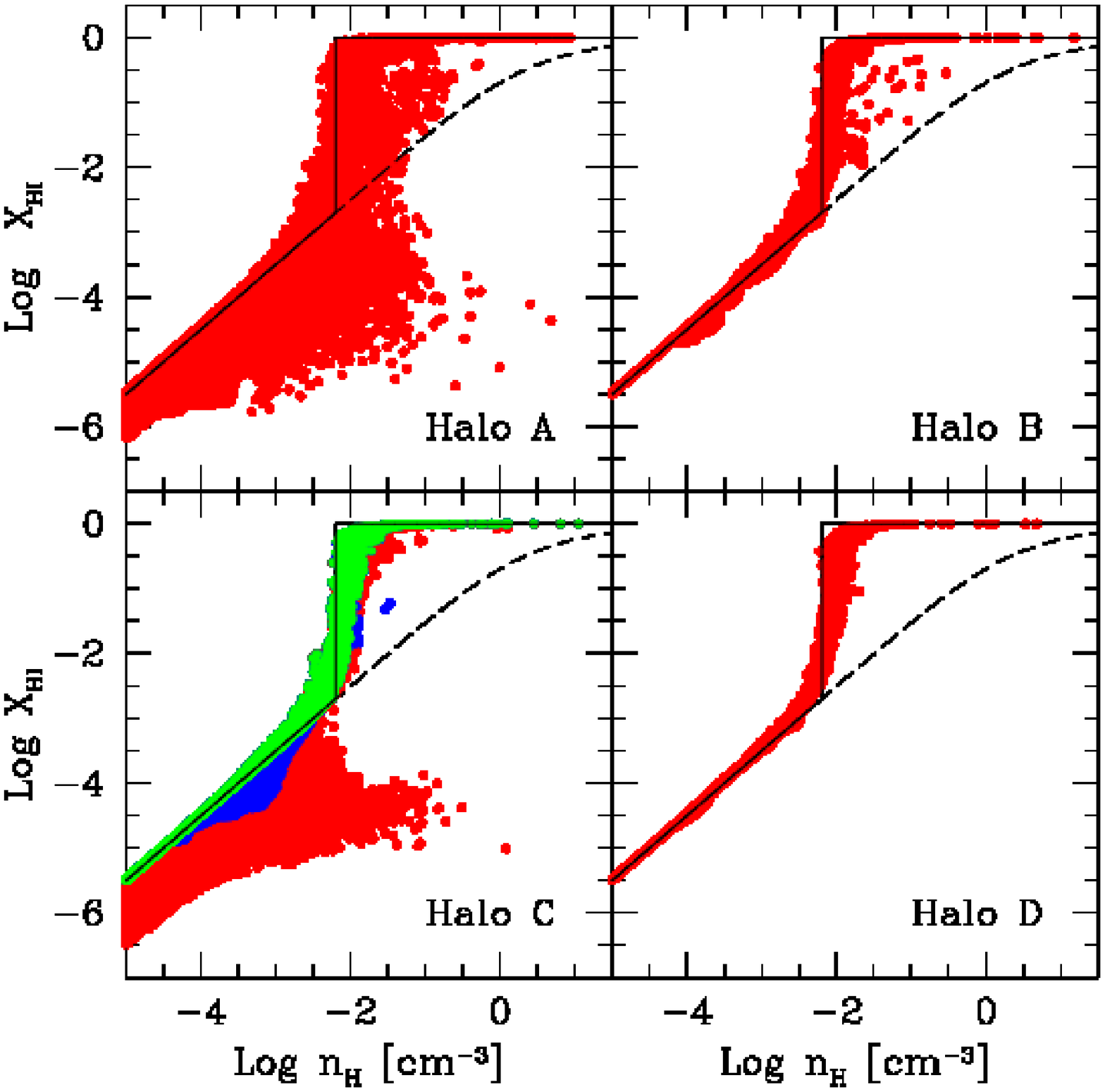}
\vspace{0.2cm}
\includegraphics[scale=0.4]{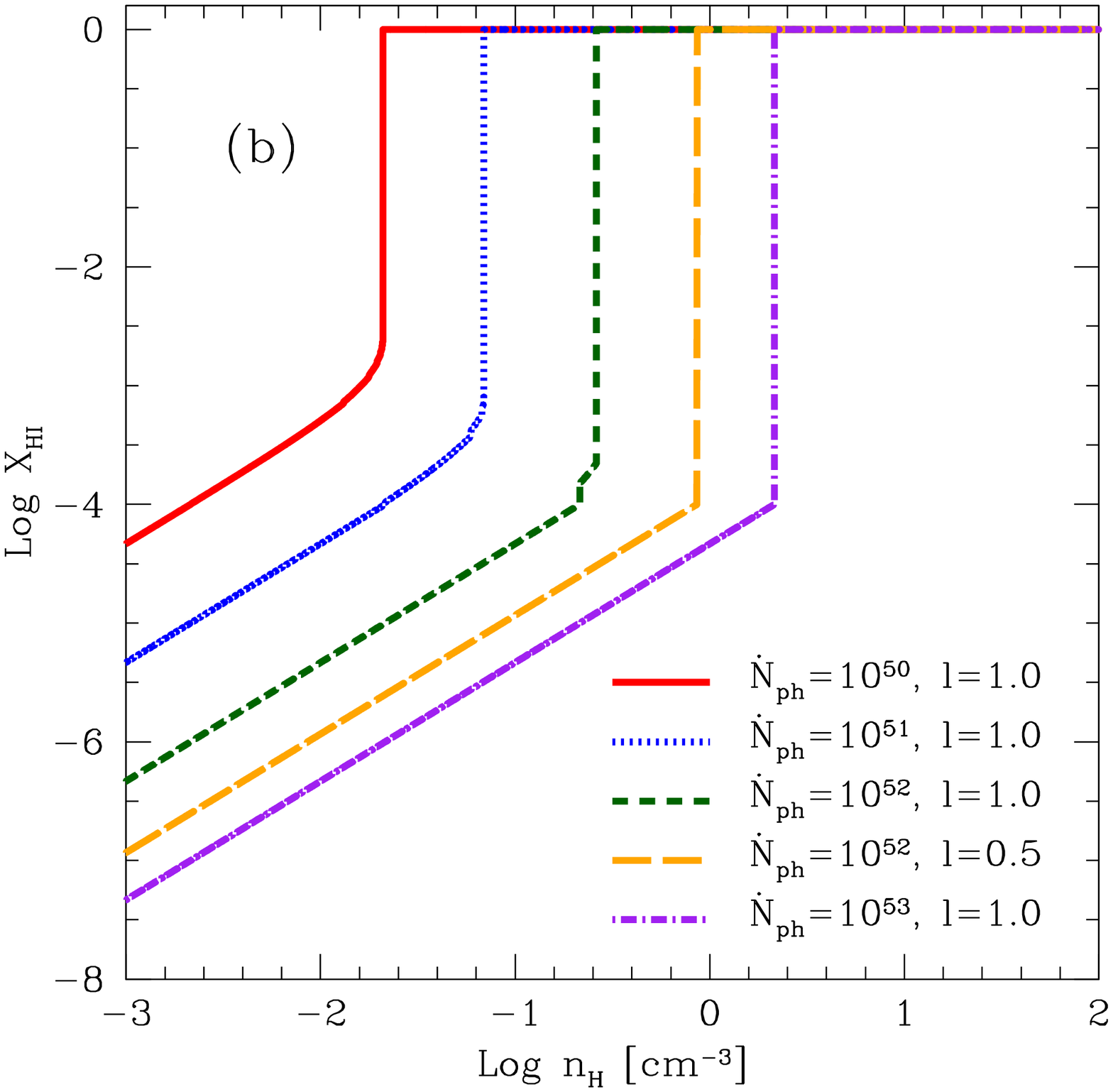}
\caption{
{\it Panel (a)} : 
Neutral fraction of hydrogen gas as a function of number density for the same halos as in Figure~\ref{fig:gammauvb}. The black dashed lines and blue solid lines indicate the equilibrium solution of ionization balance between the UVB ionization and recombination in the Optically-Thin model and the OTUV model, respectively.  The red filled circles are obtained by including direct radiative transfer calculations of UVB, stellar radiation and collisional ionization.
The blue filled circles of Halo C are the neutral fraction with UVB and collisional ionization only (i.e. without stellar radiation transfer).  The green filled circles are the neutral fraction with only UVB (i.e. without stellar radiation and collisional ionization). 
{\it Panel (b)} : Neutral fraction of gas clumps irradiated by the radiation from young star clusters at the indicated distance $\ell$ as a function of hydrogen number density. 
See text for the details of the calculation.  Different colored lines indicate different conditions as indicated in the legend, where $\dot{N}_{\rm ph}$ [s$^{-1}$] is the emissivity of ionizing photons from star clusters, and $\ell$ [kpc] is the distance between the star clusters and the gas clump. 
}
\label{fig:uvbcom}
\end{center}
\end{figure*}

Another major source of different results might be the different mass range of simulated galaxies in each simulation.  
\citet{Fumagalli11} focused on only seven star-forming galaxies with $\Mh > 5\times 10^{10} \Msun$.
On the other hand, our simulations cover a wider range of galaxy masses, and there are numerous low-mass, faint galaxies without young star clusters that contribute to the DLA cross section.  In those low-mass haloes, the local stellar radiation does not impact the ionization of gas, and Fumagalli et al.'s simulation did not include these low-mass halos.  
In our simulations, about half of the total DLA cross section is contributed by 
faint low-mass haloes where there is no young stars in our simulations. 
The DLAs in these faint haloes are unaffected by the stellar radiation.
This could explain the little effect of local stellar radiation on $\fn$ in our simulations, and more effect in Fumagalli's. 
In addition, 
Fumagalli et al.'s $\fn$ showed good agreement with observations at $\NHI \gtrsim 10^{20.3}$\,cm$^{-2}$, but underpredicted at $10^{18} \lesssim \NHI \lesssim 10^{20}$\, cm$^{-2}$.  They suggested that the reason for the underestimate might be their limited galaxy sample with $\Mh \gtrsim 5\times10^{10} \Msun$.  
For comparison, here we investigate the effect of galaxy mass limit in our simulation in Figure~\ref{fig:cdfall}. 
When we consider only the massive galaxies in haloes with $\Mh > 5\times 10^{10} \Msun$, then $\fn$ decreases for DLAs with a similar slope to the all-halo case, and $\fn$ is significantly lower and more flattened for the LLS. 
Integrating $\fn$ over the column density range of LLS, we find that the contribution from the massive galaxies to the line density of LLSs is only $\sim$20\% of the total.  Hence, we confirm that the LLSs mainly come from lower-mass galaxies, and the discrepancy of $\fn$ with observation at LLS range in \citet{Fumagalli11} is probably due to the mass limit of their galaxy sample.

In addition, we compute the $\fn$ using only the star-forming galaxies in our simulations, and find that the result is in-between the all-halo case and only massive galaxy case. 
This result suggest that LLSs mainly originate from lower mass galaxies in haloes with $\Mh < 5\times 10^{10} \Msun$ and lower star formation rate. 
We will discuss this point more quantitatively in Section~\ref{sec:typical}.

\begin{figure*}
\begin{center}
\includegraphics[width=0.95\columnwidth,angle=0] {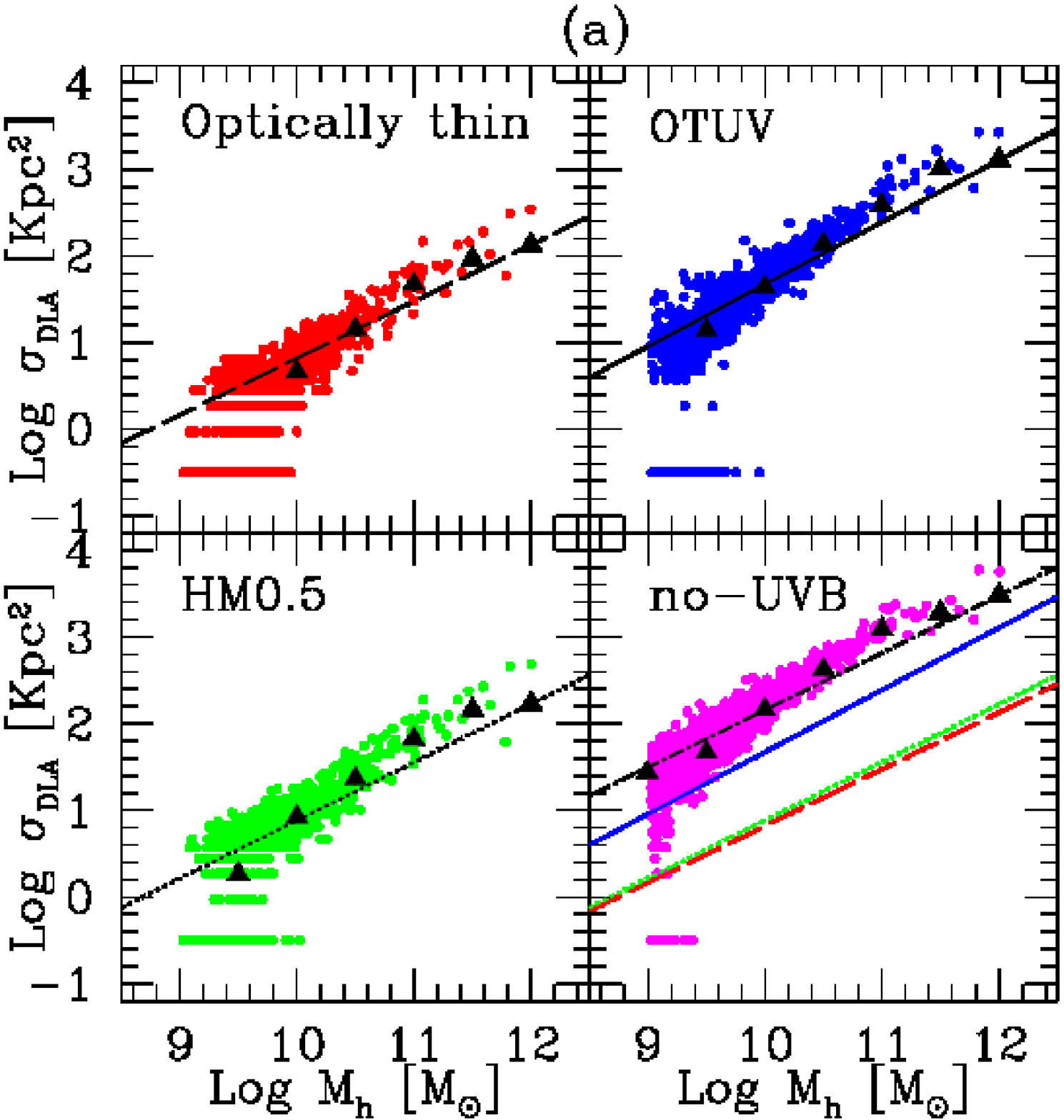}
\includegraphics[width=0.95\columnwidth,angle=0] {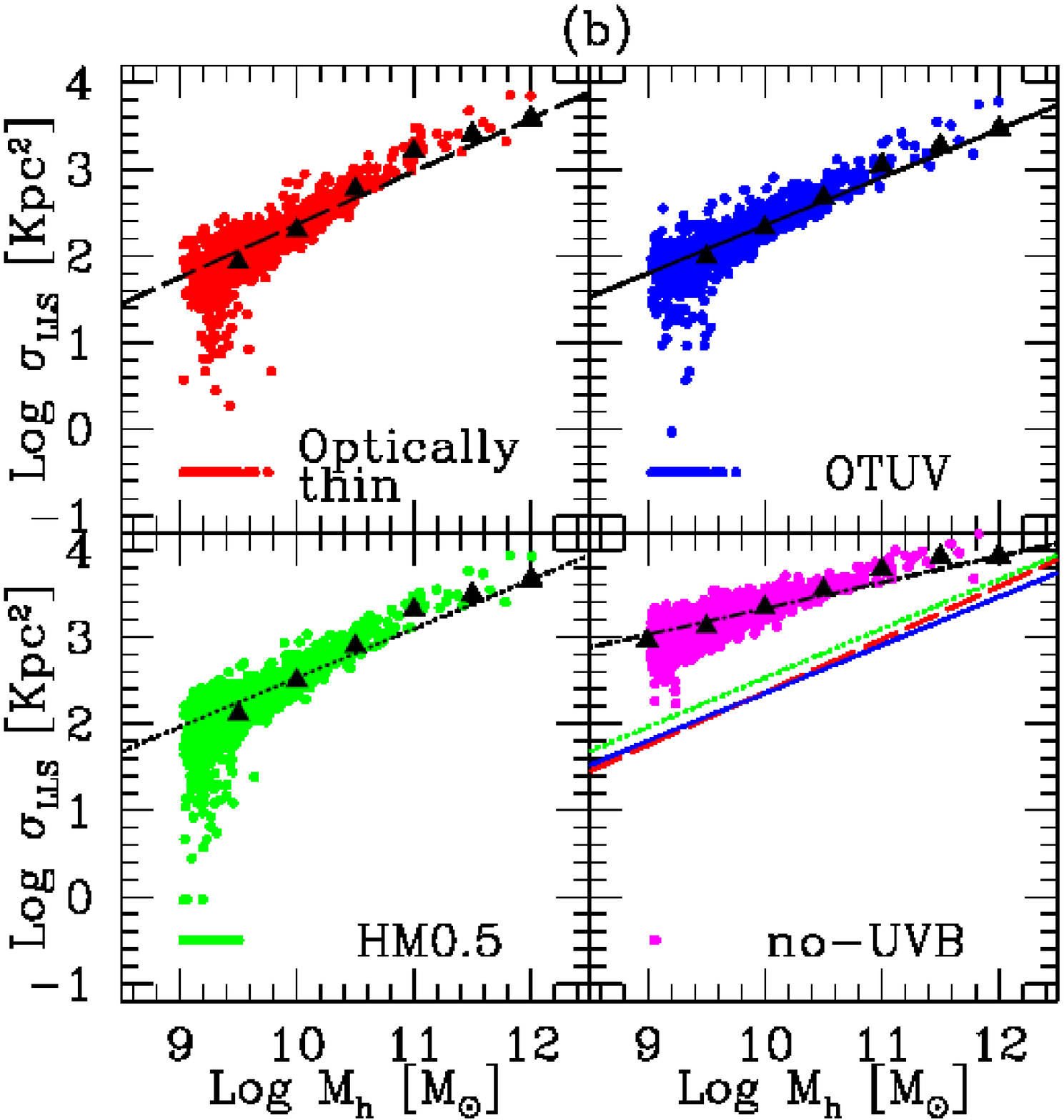}
\caption{
 Physical cross section of DLAs ({\it panel [a]}) and LLSs ({\it panel [b]}) as a function of halo mass at $z=3$.  
 Each point represents a halo with star-forming galaxies, including the stellar RT.  Filled triangles indicate the median values in each mass bin.
In the bottom right panel, the results from different runs are compared: Optically-Thin (red), OTUV (blue), HM0.5 (green) and No-UVB (magenta) runs, respectively.
}
\label{fig:sigmauv}
\end{center}
\end{figure*}

\subsubsection{Radiation field by UVB transfer}
\label{sec:UVBtransfer}

Next we perform direct radiation transfer calculation of UVB to validate the results of the OTUV run. 
We perform this calculations by irradiating UVB from the boundary surfaces of calculation boxes with $n_{\rm g}^{4}\; (\approx 10^{8-10})$ rays, and also consider the stellar radiation transfer simultaneously.

Figure~\ref{fig:gammauvb} shows the photoionization rate versus gas density in each grid cell from direct UVB transfer calculation.  Here we pick four halos randomly from the halo sample of $\Mh < 10^{11} \; \Msun$ in the OTUV run, 
and recompute the transfer of UVB directly in each cell as shown by the red points. 
The higher mass haloes are excluded from our sample, because the UVB RT calculation takes too long to process such a large grid ($>$100$^3$) with current computers. The red points in the figure are derived from the RT calculations including both the UVB and stellar radiation. The black dashed and blue solid lines are the photoionization rates by the Optically-Thin model and the OTUV model,  respectively, obtained from the equilibrium calculation. 
Note that the photoionization rate in this figure is of UVB radiation, $\Gamma_{\rm UVB}^{\gamma}$ in equation (1), not including stellar radiation $\Gamma_{\rm star}^{\gamma}$.

We find that the $\Gamma_{\rm UVB}$ has a relatively sharp transition from a high plateau at low densities to a much lower value at $\log \nh \approx -2.2$, and the self-shielding threshold density ($\nuv=6\times 10^{-3}$\,cm$^{-3}$, indicated by the blue solid line) adopted in the OTUV model correctly captures this sharp transition.  The value of $\nuv$ was determined by \citet{Nagamine10} by examining the agreement with the observed data on $\fn$, and one of the main purposes of this paper is to validate this choice of $\nuv$ through direct RT calculations as demonstrated here. \citet{Nagamine10} also gave some justification for this particular choice of $\nuv$, referring to prior radiative transfer simulations. 
In addition, the adopted threshold density is close to that of \citet{Haehnelt98}, who suggested the density threshold model based on the tight relation 
between column density and density. The region with $n_{\rm H} \ge 10^{-2}~\rm cm^{-3}$ mainly contributes to the column density of $N_{\rm HI} > 10^{17}~\rm cm^{-2}$,
which can self-shield from UVB.

Some limited number of cells at $n_{\rm HI} > 10^{-2}~\rm cm^{-3}$ for Halos A \& C show relatively high values of $\Gamma^{\gamma}$, deviating from the main photoionized branch of gas.  These are cells that are probably ionized by the local stellar radiation, and therefore allows UVB to penetrate into deeper densities. 
In some regions near star-clusters, the stellar radiation can increase the photo-ionization rate than UVB alone by orders of magnitudes,
and ionize star-forming regions. However, the impact is not so significant for a statistical quantity such as $\fn$ (Figure~\ref{fig:cdfall}).

Figure~\ref{fig:uvbcom}a shows the neutral fraction of hydrogen gas ($X_{\rm HI}$) in each cell as a function of number density for the same four halos shown in Figure~\ref{fig:gammauvb}.  
In Halos A and C, the stellar radiation boosts the ionization degree than the optically-thin equilibrium values, populating the red points below the black dashed curve. 
In Halo B and D, the young star clusters are distributed in high-density peaks, hence most ionizing photons are absorbed on the spot.  In addition, the neutral fraction $X_{\rm HI}$ in low density region is approximately fitted by $X_{\rm HI} \sim \alpha_{\rm B} \nh / \Gamma^{\gamma}_{\rm UVB}$, therefore the neutral fraction is simply proportional to the number density of hydrogen.
In addtion, 
we calculate the neutral fraction with different ionization processes (green : UVB only, magenta : UVB + collisional ionization, and red: UVB + stellar radiation + collisional ionization). 
It shows that the collision boosts the ionization degree at $n_{\rm H_I} \sim 10^{-4}~\rm cm^{-3}$ (magenta points extending below the dashed curve), and the stellar radiation boosts the ionization degree even at higher density regions, as signified by the wide distribution of red points below the dashed curve. 

We can further understand the transition from ionized to neutral state by solving for the balance between photoionization by stars and recombination in the one-zone approximation, picturing the gas clumps near young star clusters. In Figure~\ref{fig:uvbcom}b, we estimate the photoionization rate by 
$\Gamma^{\gamma}_{\rm star} = \dot{N}_{\rm ph} \sigma_{0} {\rm exp}(-\tau) / 4\pi l^{2}$,
where $\dot{N}_{\rm ph}$ is the emissivity of ionizing photons from star clusters, $ \sigma_{0}$ is the ionization cross section at the Lyman limit, $l$ is the distance between star clusters and gas clumps.
The optical depth $\tau$ is roughly estimated by $\tau = \sigma_{0} n_{\rm HI} l$, where $n_{\rm HI}$ is the number density of neutral hydrogen of the gas clump.
The $\dot{N}_{\rm ph}$ of a star cluster in our simulations can be $\sim 10^{50-52}$\,s$^{-1}$, and the distance $l$ can be proper $\sim 0.5 - 1.0$\,kpc, corresponding to the cell size of RT calculation in our simulation. 
In Figure~\ref{fig:uvbcom}b,  the neutral fraction steeply rises at the density where $\tau$ becomes $\sim 1$. 
With increasing photon emissivity, the threshold density moves to a higher density as the radiation is able to sink into higher densities, and the ionization degree at lower densities become stronger.  With the above parameter values, we find that the star clusters cannot ionize gas clumps with $n_{\rm HI} \gg 1\; \rm cm^{-3}$.  In our simulations, the gas density near star clusters can be $> 1$\,cm$^{-3}$ in massive halos and block the ionizing radiation from stars.
From these analyses of UVB radiation transfer, we confirm that the stellar radiation does not strongly affect the neutral fraction of high-density gas at $\nh > \nuv$, and the UVB treatment of the OTUV run is reasonable.

\subsection{Cross Sections of DLAs and LLSs}
\label{sec:sigma}

Once the \HI\ column density of each cell in the projected plane is obtained, we can estimate the DLA cross-section of each halo by counting the number of grid-cells that exceed $\NHI = 2 \times 10^{20}$\,cm$^{-2}$ and multiplying by the unit area $\left(\epsilon/(1+z)\right)^{\rm 2}$ of the grid-cell.  In this paper, we discuss the cross section in physical units.

First, we examine the effect of different UVB models on the cross section.
Figure~\ref{fig:sigmauv}a,b show $\sigmadla$ and $\sigmalls$ of each halo as a function of halo mass for different UVB runs. 
Each point in the figure represent a halo with star-forming galaxies, including the effect of stellar RT. 
Each run has $\sim 2000$ galaxies in the simulation box at $z = 3$, and about 200 of them have young star clusters ($t_{\rm age} \le 10^{7}~\rm yr$) that can affect the hydrogen ionization (hereafter we call these galaxies ``star-forming galaxies'').  Here, we select the galaxies with young stars from each run, and compute the RT of local stellar sources. 

We find that the cross section increases with increasing halo mass for both DLAs and LLSs in all UVB models. For a quantitative discussion, we plot the median value in each bin with a triangle symbol, and then fit a power law $\sigmadla  \propto \Mh^{\alpha}$, assuming a functional form of
\begin{equation}
\log \sigma = \alpha (\log \Mh - 12) + \beta, 
\label{eq:sigma}
\end{equation}
where $\Mh$ is the dark matter halo mass in units of solar masses, and the values of slope '$\alpha$' and normalization '$\beta$' are determined by least-square fitting.  We use the cross section at $\Mh = 10^{12} \Msun$ as the anchor point of $\beta$.  The best-fit values of $\alpha$ and $\beta$ of all runs are summarized in Table~\ref{table:alpha}.

The bottom right panels of Figure~\ref{fig:sigmauv}a,b compare the results of different UVB models. 
For both DLAs and LLSs, the cross section is the greatest for the No-UVB run, and it drops dramatically once the UVB is turned on. The difference between the Optically-Thin run and the HM0.5 run is small, suggesting that the gas is ionized to relatively high densities under the optically-thin approximation, irrespective of the specific strength of the UVB. 
For DLAs, the OTUV run has much higher cross sections than in the HM0.5 run, because the high-density clouds with $n>\nuv$ are the main contributor to $\sigmadla$, and the OTUV model increases the neutral fraction of such gas over the optically-thin approximation. 
In fact, the value of $\beta$ for the OTUV run is much higher than that of HM0.5 run, but the slope is similar in all runs with $\alpha \approx 0.65-0.72$ for DLAs. 
This suggests that $\sigmadla$ is affected proportionally at all halo mass range by the different UVB models. 
Obviously the result of the Optically-Thin run is very similar to that of \citet{Nagamine04f}, as they both used the optically-thin approximation in similar simulations (the main difference is the galactic wind model). 

\begin{table}
\begin{center}
\begin{tabular}{ccccccc}
\hline
Run &  $\alpha$(DLA) & $\alpha$ (LLS) & $\beta$ (DLA) & $\beta$ (LLS)  \\
\hline
Optically-Thin     & 0.65   & 0.61 & 2.12 & 3.59  \\
HM0.5   & 0.67 & 0.56  & 2.23 & 3.66  \cr
OTUV   & 0.70 & 0.53  & 3.06 & 3.41  \cr
OTUV (no-star)       & 0.72   & 0.55 & 3.11 & 3.46  \\
no-UVB   & 0.66 & 0.30  & 3.48 & 3.94  \cr
\hline
\end{tabular}
\caption{
Best-fit parameters of the fitting function (Eq.\,[\ref{eq:sigma}]) for $\sigmadla$ and $\sigmalls$ for different UVB models.
The value of $\alpha$ is the slope of the power law, and $\beta$ is the normalization point at $\Mh = 10^{12}\Msun$. 
}
\label{table:alpha}
\end{center}
\end{table}

The biggest difference between $\sigmadla$ and $\sigmalls$ is that the OTUV run (blue line) has almost the same $\sigmalls$ as the Optically-Thin run (red line), whereas $\sigmadla$ are very different in the two runs.  This implies that the high-density gas with $n>\nuv$ is not responsible for the LLS.  The HM0.5 run has a slightly higher $\sigmalls$ than the Optically-Thin run as expected. 
We also find that the slope $\alpha$ is systematically shallower for the LLSs compared to that of DLAs, and the normalization $\beta$ is higher for the LLSs.  This is expected, because LLSs are more extended in the outskirts of halos. 

If we look at the distribution of data points in Figure~\ref{fig:sigmauv}a,b more carefully, one can see that there is a slight curvature to the median points, and a straight power-law does not necessarily provide a perfect fit.  
At low halo masses, the median points fall below the power-law fit. At the intermediate to high masses, the median points are slightly higher than the power-law fit, and the cross section seems to flatten out at the highest masses at $\Mh \sim 10^{12} \Msun$.  A similar flattening has been observed in other works \citep{Pontzen08, Fumagalli11}, as we will discuss in the next section in more detail.

\begin{figure}
\begin{center}
\includegraphics[scale=0.42] {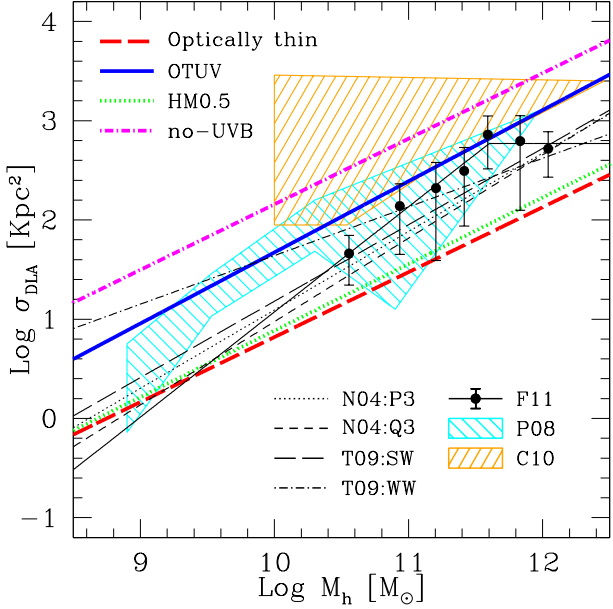}
\caption{
Comparison of DLA cross section with other theoretical works.
The black dotted and dashed lines are the results of P3 and Q3 models, respectively, by \citet{Nagamine04g}. 
The black long-dashed and dot-dashed lines are the strong wind (SW) and weak wind (WW) runs, respectively, in the GADGET SPH simulations of \citet{Tescari09}.
The black filled circles are the cross sections of seven simulated galaxies in the AMR simulation of \citet{Fumagalli11}.
The cyan and orange shade enclose the data points from the simulations of  \citet[][Gasoline SPH simulation]{Pontzen08} and \citet[][Enzo AMR simulation]{Cen10}, respectively.
}
\label{fig:sigma_com_model}
\end{center}
\end{figure}

\begin{figure}
\begin{center}
\includegraphics[scale=0.4]{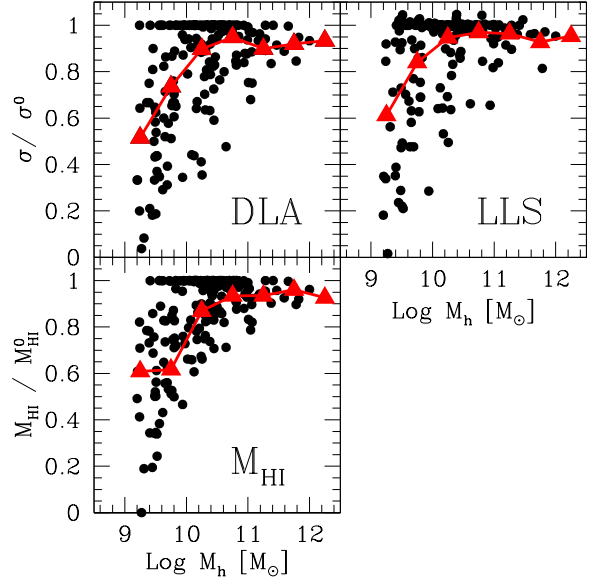}
\caption{
{\it Upper panels} : 
The ratio of cross sections of DLAs and LLSs in each halo with ($\sigma$) and without ($\sigma^{0}$) stellar RT: 
i.e., $\sigma / \sigma^{0}$. 
{\it Lower panel} : The mass ratio of neutral hydrogen in each halo with ($M_{\rm HI}$) and without ($M^{0}_{\rm HI}$) stellar radiation,
i.e., $M_{\rm HI} / M^{0}_{\rm HI}$. 
The red points are the mean value in each bin with the bin size of 0.5 dex.
}
\label{fig:dlacom}
\end{center}
\end{figure}

\subsubsection{Comparison with Other Simulation Results}

In Figure~\ref{fig:sigma_com_model}, we compare the results of DLA cross section from several authors using different simulations \citep{Nagamine04f, Pontzen08, Tescari09, Cen10, Fumagalli11}. 
Although there are small differences in the slope and normalization, most of the results are bracketed by our no-UVB run and Optically-Thin run. 
There are significant overlap between the results of our OTUV run, \citet{Pontzen08}, \citet{Fumagalli11}, and the WW (weak wind) model of \citet{Tescari09}, which is very encouraging. 

\citet{Pontzen08} considered a crude self-shielding model, so it is natural that their result is close to the OTUV run. 
Although \citet{Tescari09} also set the self-shielding threshold density criteria above which UVB cannot penetrate,
their $\nuv$ is about one magnitude higher than our value.
Hence, their SW (strong wind) model should give lower $\sigmadla$ than our OTUV run.
The SW and WW model have constant wind particle velocities of $600$ and 100\,km\,s$^{-1}$ as a stellar feedback, respectively.
The difference between the SW and WW model of \citet{Tescari09} is consistent with the results of \citet{Nagamine04f}.

Using an Enzo AMR simulation, \citet{Cen10} reported a much higher $\sigmadla$ than other simulations for halos with $\Mh=10^{10}-10^{11} \Msun$, and the reason for this  difference is not very clear.  Some of the possibilities are: 1) their star formation model is considerably inefficient in converting the neutral gas into stars, 2) their self-shielding treatment is allowing too much neutral gas in massive halos, and/or 3) their stellar feedback is inefficient and does not heat up the ambient neutral gas.  They used only two zoom-in simulations (a high-density cluster region and a low-density void region) to bracket the mean density result, therefore they only bracketed the observed $\fn$ without reproducing the normalization of $\fn$.  It appears that his result is more consistent with that of our No-UV run rather than the OTUV run. 

We find that the Fumagalli's fit to the $\sigmadla$ for low-mass halos is much steeper than other simulations, which could be due to a small galaxy sample in their zoom-in simulations (only seven galaxies).  We discussed the effect of limited galaxy sample already in \S~\ref{sec:fn} and Figure~\ref{fig:cdfall}. 

\begin{figure*}
\begin{center}
\includegraphics[scale=0.8]{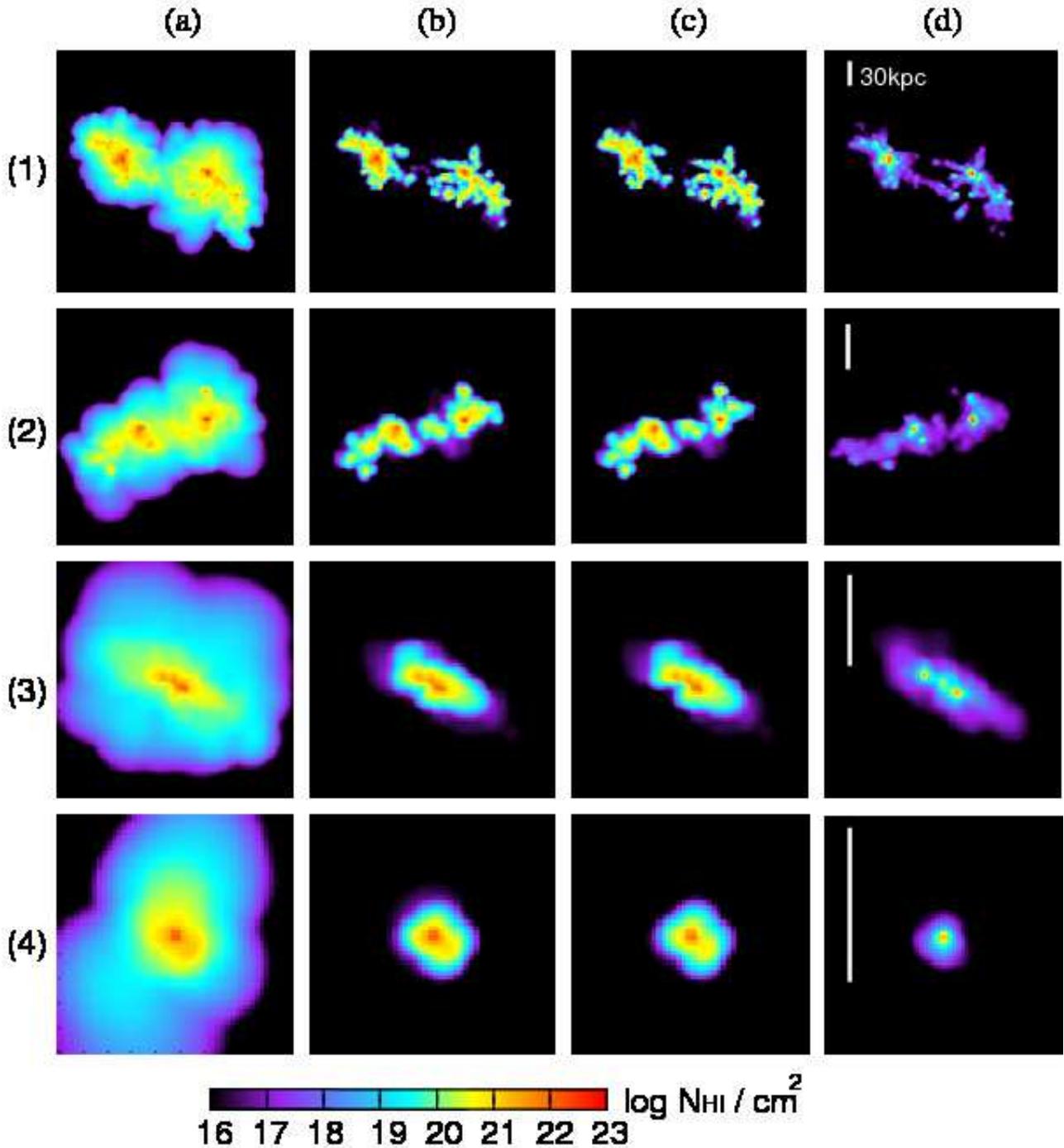}
\caption{
Two dimensional map of $\NHI$ for four halos in the OTUV run at $z=3$.
Each column shows different treatment of radiation:
column ($a$): no-ionization;  ($b$): UVB $+$ collisional ionization, ($c$): UVB $+$ star $+$ collisional ionization.
The column ($d$) is the same as column ($c$), but for the same halo in the Optically-Thin run.
Each row corresponds to one halo with a panel size $\ell_{\rm phys}$ in physical units: 
Halo (1): $\Mh = 6.7 \times 10^{11} ~{\rm \Msun}, \ell_{\rm phys} = 300~ {\rm kpc}$; 
Halo (2) $\Mh = 1 \times 10^{11} ~{\rm \Msun}, \ell_{\rm phys} = 159~ {\rm kpc}$; 
Halo (3) $\Mh = 2.6\times 10^{10} ~{\rm \Msun}, \ell_{\rm phys} = 80 ~{\rm kpc}$; and 
Halo (4) $\Mh = 3.3 \times 10^{9} ~{\rm \Msun}, \ell_{\rm phys} = 47~ {\rm kpc}$.
The white vertical tick mark in column ($d$) indicates a scale of proper 30 kpc. 
}
\label{fig:column}
\end{center}
\end{figure*}

\subsubsection{Effect of Stellar RT on \HI\ Cross Sections}
\label{sec:stellarRT}

The upper panels of Figure~\ref{fig:dlacom} compares the ratio of DLA and LLS cross section with and without stellar radiation.
The data points are only for those halos that contain star-forming galaxies in the OTUV run at $z=3$.
We see that both $\sigmadla$ and $\sigmalls$ of lower mass haloes are reduced by the stellar RT, but the higher mass haloes are not affected very much. 
The slope $\alpha$ and normalization $\beta$ for the power-law fit are very similar in the two cases of with and without the stellar RT. 
Since young stars distribute near high-density gas clumps, most of ionizing photons from stars are blocked by the gas clumps, and the region behind high-density gas clumps are not irradiated by the stellar radiation. 
As a result, the stellar radiation cannot propagate over long distances in the halo, and the stellar radiation does not affect the \HI\ cross section very much in high-mass halos. 
In addition, although relative mass of neutral hydrogen gas to stellar mass somewhat decreases as increasing halo mass,
the amount of dense gas clouds around star-forming regions increases with halo mass by frequent minor merging processes \citep[e.g., Figure~1~in][]{Yajima11}.
The photon absorption process and the distribution of stars and gas are determined by complicated nonlinear process,
which can be traced by cosmological hydrodynamics and radiative transfer simulations.

The lower panel of Figure~\ref{fig:dlacom}
shows the ratio of \HI\ mass in each halo with ($M_{\rm HI}$) and 
without ($M^{0}_{\rm HI}$) the stellar RT, i.e., $M_{\rm HI} / M^{0}_{\rm HI}$. 
It shows that the lower mass halos lose more \HI\ masses than the higher mass halos, consistently with Figure~\ref{fig:dlacom}. 
The \HI\ mass is reduced by about 30 per cent by the stellar RT for halos with $\Mh < 10^{10}\;\Msun$, 
while it is only about 7 per cent reduction for higher mass halos with $\Mh > 10^{11}\;\Msun$. 

Some analytical works discussed the relative strength of local stellar radiation to UVB under the assumption of 
simple gas distribution \citep{Miralda05, Schaye06}.
\citet{Miralda05} showed that the effect of stellar radiation is negligible compared to UVB for LLSs, and that it can be comparable for DLAs as an upper limit.
On the other hand, the model by \citet{Schaye06} indicated that the local stellar radiation can be dominant even for LLSs, as well as for DLAs.
However, for inhomogeneous gas distribution, the effect of stellar radiation strongly depends on the structure of gas around stars, and it requires RT simulations for proper assessment.  The above analytical works did not discuss the direct impact of local stellar radiation on DLA/LLS cross sections and $\fn$, and focused only on the {\it relative} strength between local stellar radiation and UVB. 
In our simulations, the ionizing flux from local stars can be much higher than the UVB similarly to the previous analytic estimates.
However, the neutral clouds survive from ionization by self-shielding and shadowing effect \citep[See Fig.1 of][]{Yajima11}. 
As a result, the local stellar radiation cannot change the ionization structure in galaxies very much (particularly in massive galaxies),  and hence the \HI\ cross section does not change largely. 
On the other hand, the UVB enter from outside the halo into low-density region, therefore they can irradiate over greater volume.


\subsection{Structure of DLAs and LLSs}

As we see in Figure~\ref{fig:cdfall}, the UVB can drastically change the rate-of-incidence of DLAs and LLSs.
In this subsection, we investigate how the UVB and stellar radiation change the spatial distribution of $\NHI$. 
Figure~\ref{fig:column} shows the two dimensional map of $\NHI$ distribution for four halos in the OTUV run at $z=3$. 
The column ($a$) is without any ionization by UVB and stellar radiation, corresponding to the No-UV run. 
The column ($b$) is with the UVB RT and collisional ionization, but no stellar RT. 
The column ($c$) is with the stellar RT in addition. 
The column ($d$) is the same as column ($c$), but for the same halo in the Optically-Thin run. 

Comparing columns ($a$) and ($b$), we see that the UVB changes the $\NHI$ distribution drastically. 
Much of the LLSs in the outer region are ionized by the UVB and becomes lower than $\NHI = 10^{16}$\,cm$^{-2}$. 
The higher columns with $\NHI \gtrsim 10^{20.5}$ survive from photo-ionization by UVB because of the high recombination rate in high density regions, but the DLA cross section is significantly reduced. 

The DLAs correspond to the yellow and red regions in Figure~\ref{fig:column}.
The DLA distribution does not change very much from column ($b$) to ($c$), and it is not affected by the stellar RT very much as we discussed in \S~\ref{sec:stellarRT}. 

Comparison of columns ($c$) and ($d$) reveals the effect of optically-thin approximation of UVB in the Optically-Thin run relative to the self-shielding effect in the OTUV run. 
It shows that the UVB photo-ionizes the gas too deeply in the optically-thin approximation, and the self-shielding model increases the DLA cross section significantly. 
From the discussion in \S~\ref{sec:fn}, we argue that column ($c$) is more realistic than column ($d$) based on the comparison of $\fn$. 

In the OTUV run (column [$c$]), DLAs are extended over $\sim 10$\,kpc scales in all halos, which is a stark contrast to the very compact appearance in the Optically-Thin run (column [$d$]), and the LLSs are surrounding DLAs on $\sim 30-60$\,kpc scales. 
\citet{Nagamine04f} showed the DLA distribution similar to the column ($d$) in their SPH simulations with optically-thin approximation and argued that DLA distribution is very clumpy.  However based on the present work, it seems that DLAs are more broadly and smoothly distributed around LBGs at $z=3$ on $\sim 10$\,kpc scales. 
The highest column density systems with $\NHI > 10^{22}$\,cm$^{-2}$ are preserved well even in the Optically-Thin run, as we saw in Figure~\ref{fig:cdfall}.

\begin{figure}
\begin{center}
\includegraphics[scale=0.45]{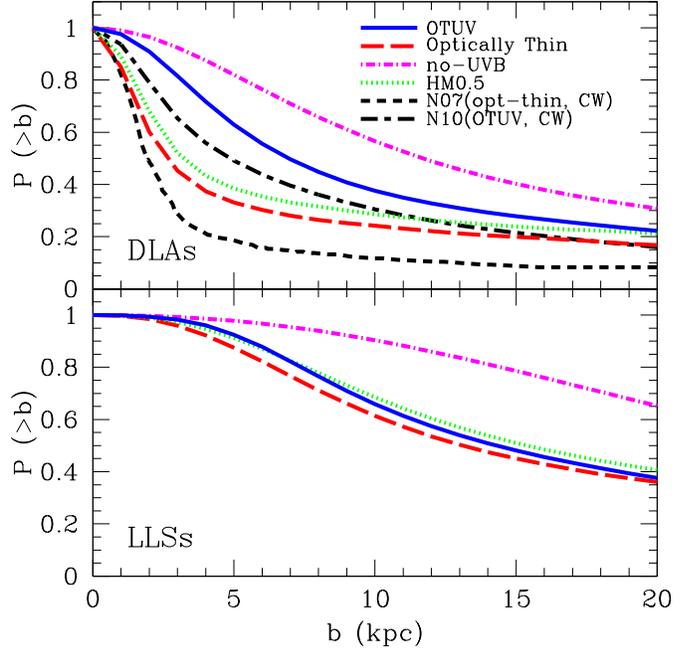}
\caption{
Cumulative probability distribution function (PDF) of DLAs and LLSs as a function of separation distance (i.e. impact parameter) 
between DLAs (LLSs) and center-of-mass of the nearest galaxy in projected 2D plane in physical unit.
The colored lines are our simulation results in this work.
The black dash and dot-dash lines are PDFs with the constant-velocity (CW) wind model + optically-thin UVB approximation \citep[P3 run in][]{Nagamine07} and the CW+OTUV run \citep{Nagamine10}, respectively.
}
\label{fig:PDF}
\end{center}
\end{figure}

To quantify the relative distribution of DLAs, LLSs and galaxies in different UVB models, 
we compute the impact parameter distribution of DLAs/LLSs relative to galaxies in each run.
Figure~\ref{fig:PDF} shows the cumulative probability distribution function (PDF) of DLAs and LLSs 
as a function of separation distance (i.e. impact parameter) between DLAs (LLSs) and the center-of-mass of nearest galaxy in projected 2D plane in physical unit. 
The colored lines are from the simulations used in this paper.
The OTUV run shows a much shallower PDF than the Optically-Thin run.
About fifty per cent of DLAs in the OTUV run distribute within $\sim 7$~kpc from galaxy center,
while those in the Optically-Thin run are concentrated within $\sim 3$~kpc.
This is because the UVB cannot penetrate deeper into the galaxy center due to self-shielding, and 
hydrogen gas near galaxy center can survive as DLAs.
Of course, no-UVB run has the widest PDF due to the overproduction of DLAs as noted earlier in this paper. 

In addition, we compare the results with different galactic wind models.
As described in Section 2.1, 
we use the "Multi-component Variable Velocity" (MVV) wind model of \citet{Choi10} 
in the current simulations.
Some previous works used the constant-velocity wind (CW) model.
The black dashed and solid lines are the PDFs from the CW model + optically-thin UVB approximation \citep[P3 run in][]{Nagamine07} and CW+OTUV run \citep{Nagamine10}, respectively.
We find that the DLAs in the CW model are concentrated near the galaxy center more than the MVV model.
In the CW model, the lower mass galaxies have higher wind velocities than in the MVV run \citep{Choi10}, 
therefore the DLAs abundance is somewhat suppressed.
In such a situation, only high-density gas around galaxy center can survive as DLAs, resulting in a steeper PDF. 
In the case of LLSs, the PDFs are in general much wider than those of DLAs'. 


\subsection{Stellar Mass and Metallicity Distribution}

\begin{figure*}
\begin{center}
\includegraphics[scale=0.8]{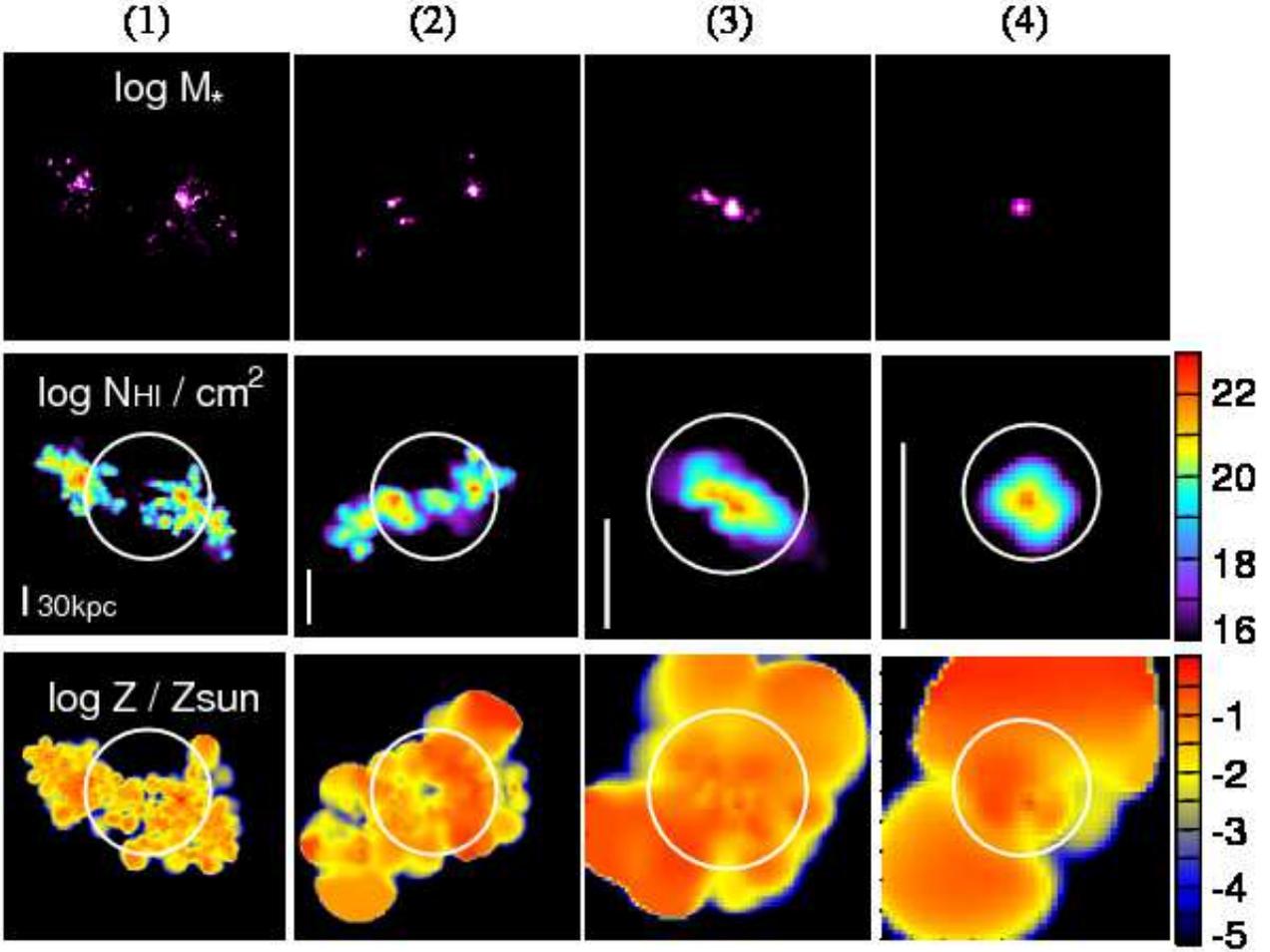}
\caption{
Two dimensional map of projected stellar mass (top row), $\NHI$ (middle row), and metallicity (bottom row) for the same four halos shown in Figure~\ref{fig:column} in the OTUV run.
Each column with the number index corresponds to the same halo in each row of Figure~\ref{fig:column}, and the panel sizes are the same as in Figure~\ref{fig:column}. 
The color scheme of stellar mass is normalized in each panel. 
The middle row is the same as the column (c) in Figure~\ref{fig:column}, but we show it again for comparison. 
The white bar in the middle row shows the proper 30 kpc.  Note the different panel sizes for each column. 
The metallicity map is obtained by summing up all metal mass along each line-of-sight and dividing by the total gas 
mass in the same sight-line.  The white circles indicate the virial radius of each halo: physical 68, 37, 23 and 11 kpc for halos (1)-(4), respectively. 
The virial radius is estimated by $R_{\rm vir} \simeq 144 (1 + z_{\rm vir})^{-1} (\Mh / 10^{11} \Msun)^{1/3}$\,kpc \citep[e.g.,][]{Mo02},
and here we assumed $z_{\rm vir} = 3$. 
}
\label{fig:mtl}
\end{center}
\end{figure*}

We show the projected maps of stellar mass, $\NHI$, and metallicity in Figure~\ref{fig:mtl}.
We find a strong correlation between high-$\NHI$ systems and stellar mass distribution as expected. 
The metallicity map is obtained by summing up all metal mass along each line-of-sight and dividing by the total gas mass in the same sight-line. 
The high density region is effectively enriched by the metals created by SNe.
The massive halos (1) and (2) show high metallicities around the high density regions and DLAs.

Galactic wind can also carry high-metallicity gas outside the galaxies. 
All bottom panels of Figure~\ref{fig:mtl} show that the near-solar metallicity gas is carried well outside of the stellar distribution, up to $\sim 30-50$\,kpc away from stars.  
The effect of wind appears stronger in halos (3) and (4), but this is simply due to the smaller physical scale of the panels for halos (3) \& (4) compared to (1) \& (2).  Careful examination shows that halos (1) \& (2) also show a similar degree of circumgalactic enrichment as halos (3) \& (4). 
For example, for halo (1) there is a bubble of sub-$\Zsun$ gas extending toward north-east direction, about 60\,kpc away from the center of galaxy on the right-hand-side.  For halo (2), a similar feature can be seen for a near-$\Zsun$ gas extending out towards north-east direction, again about 60\,kpc away from the center of galaxy on the right-hand-side.  For halo (3), there is a large bubble of near-$\Zsun$ gas extending towards south-west direction, about 30\,kpc away from the stars. For halo (4), a similar feature can be seen towards north direction. 
For halos (1)$-$(3), notice that the direction of wind propagation is nearly perpendicular to the chain of galaxy distribution. 

Our MVV wind model
reproduces the observed scaling between wind velocity and galaxy SFR: $v_w \propto SFR^{1/3} \propto M_\star^{1/3}$, which allows greater wind velocity for more massive galaxies.  This is compensated by the deeper potential well of more massive galaxies, therefore the enriched metals reach similar distances for all halos as we discussed above. 

The bottom two rows of Figure~\ref{fig:mtl} also show the virial radius of each halo in white circles. 
One can see that the DLAs are about the same or slightly smaller than the virial radius, whereas the metal enrichment clearly goes beyond the virial radius.  For halos (1) \& (2), there are two major galaxies in these massive halos, and each of the galaxies is a host of DLAs that extends to the scale of virial radius. 

\subsection{Cumulative DLA Rate of Incidence}
\label{sec:abundance}

Using the fitting function of DLA cross section derived in \S~\ref{sec:sigma},
we are now able to estimate the cumulative number of DLAs per unit redshift.
However, there is the mass-resolution limit in a simulation, and we cannot resolve the lower mass haloes that may be DLAs host. 
It may cause the underestimation of the DLA abundance. In addition, it is hard to produce very massive halos in a limited simulation volume.
To overcome this resolution limit, the DLA abundance has been estimated by combining a theoretical fit to the dark matter halo mass function and the derived relationship between DLA cross-section and halo mass
\citep{Gardner97a, Gardner97b, Gardner01, Nagamine04a, Nagamine07, Tescari09}.
Here, we follow the same method. 
The cumulative number of DLAs is estimated by 
\begin{equation}
\frac{dN_{\rm DLA}}{dz} (> M, z) = \frac{dr}{dz} \int^{\infty}_{M} n_{\rm h}(M^{'}, z) \sigma_{\rm DLA}(M^{'}, z) dM^{'},
\end{equation}
where $ n_{\rm h}(M^{'}, z)$ is the dark matter halo mass function, for which we use the function derived in \citet{Tinker08}, and $dr / dz = c / H_{0} \sqrt{\Omega_{\rm m}(1 + z)^{3} + \Omega_{\rm \Lambda}}$.

\begin{figure}
\begin{center}
\includegraphics[scale=0.4]{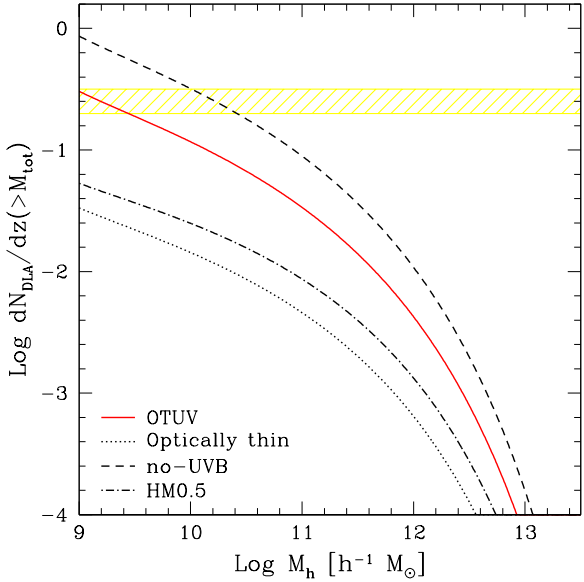}
\caption{
Cumulative abundance of DLAs per unit redshift as a function of halo mass 
for the Optically-Thin (dotted line), HM0.5 (dot-dash line), OTUV (red solid line) and No-UVB (dash line) runs at $z=3$.
The yellow shaded region shows the observed cumulative DLA abundance of 
\citet{Prochaska05} from SDSS data.
}
\label{fig:dndz}
\end{center}
\end{figure}

Figure~\ref{fig:dndz} shows the cumulative DLA number as a function halo mass for different UVB models. 
The yellow shaded region is the observational estimate $\log (dn/dz) = - 0.6 \pm 0.1$ at $z=3$ from SDSS data of \citet{Prochaska05}.
We find that the cumulative abundance largely depends on the UVB models.
The OTUV run agrees well with the observed data
, if the cross section steeply drops off at $\Mh < 10^{9}\,\Msun$. 
In this work, we set the lower limit of halo mass to $10^{9}\,\Msun$ for the above integration, 
because the DLA cross section steeply declines at halo masses of $10^{8.5}-10^9 \Msun$ 
\citep{Nagamine04g, Pontzen08}.
However, there is some ambiguity in the DLA formation in lower mass haloes, and we would overpredict the abundance 
if we were to integrate down to $10^{8.5}\,\Msun$ for the above integration. 

On the other hand, the No-UVB run clearly overproduces DLAs.
The Optically-Thin and HM0.5 runs underpredict the observation by some factor.
The result of the Optically-Thin run is slightly lower than that of \citet{Nagamine04f, Nagamine07}.
The new MVV wind model used in this work shows the smaller $\sigmadla$ at higher halo mass.
In addition, the halo number density by \citet{Tinker08} with WMAP 7-year parameters is
slightly lower than the mass function used in their paper.
Hence, it results in smaller DLA abundance even with the same optically-thin UVB prescription. 
Although our $\sigmadla$ in the OTUV run is somewhat larger than that of SW run in \citet{Tescari09},
our result still agrees with the observation data well.

\subsection{Typical physical quantities of DLA host galaxies}
\label{sec:typical}

\begin{figure}
\begin{center}
\includegraphics[scale=0.43]{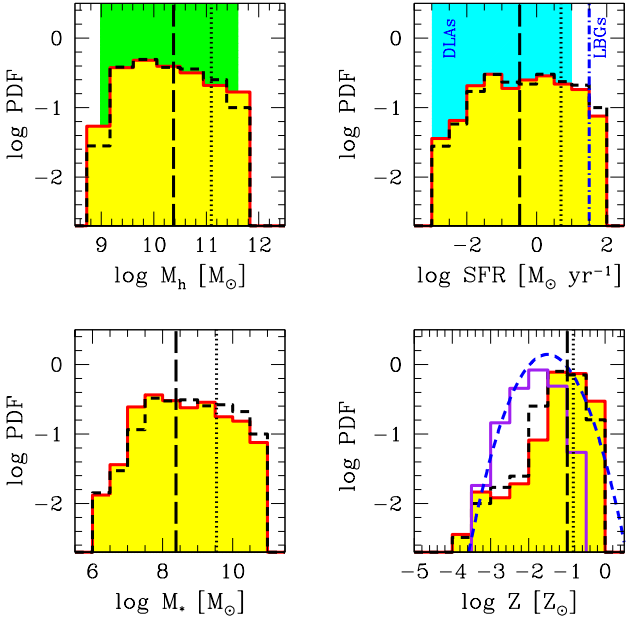}
\caption{
The PDF of DLA cross section in the OTUV run at $z=3$ as a function of various physical quantities: halo mass, SFR, stellar mass, and metallicity, from top-left to lower right, respectively.
The dashed histograms are PDFs in the Optically-Thin run.
The vertical dashed and dotted lines are the median and mean of PDF, respectively.
The green shaded region is the halo mass range of DLAs host galaxies suggested by the clustering analysis \citep{Cooke06}.
The cyan shaded region is an approximate range of SFR for DLAs derived by the $\rm C_{[II]}$ cooling technique
\citep{Wolfe03}.
The vertical, blue dot-dashed line in the upper right panel is the SFR of $L^*$--LBGs at $z=3$ \citep{Reddy09}.
The purple histogram in the bottom right panel shows the PDF of observed DLA metallicities at $2 \le z \le 4$ by  \citet{Prochaska07a}.
The blue dashed curve is the Gaussian fit to the metallicity of 207 DLAs at $1.5 \le z \le 5$ \citep{Rafelski12}.
}
\label{fig:crossdistdla}
\end{center}
\end{figure}

In this section, we examine the typical physical quantities of DLA host galaxies in the OTUV run.
Figure~\ref{fig:crossdistdla} shows the $\sigmadla$-weighted probability distribution function (PDF) as a function of certain physical quantity.  For example, PDF\,$\propto d\sigmadla / d\log M_h$ for the top left panel. 
The histograms are normalized such that the area under the histogram is equal to unity. 
The dashed and dotted lines indicate the median and mean values weighted by the cross section, respectively. 
We find the median halo mass by rank ordering the halo mass according to the $x$-axis quantity, and determining the halo at which the stacked cross section being half of the total.
We see that the mean values are biased towards higher values and not representative of the entire distribution, except for the bottom right panel where the distribution is strongly peaked at a certain metallicity.  Therefore median value is the better indicator of typical property of DLA hosts. 

The upper left panel of Figure~\ref{fig:crossdistdla} shows that the median halo mass of DLA hosts is $\Mh = 2.4 \times 10^{10} \Msun$, and about half of the total $\sigmadla$ is contributed by the halos with $\Mh = 6.6 \times 10^{9} - 1.1 \times 10^{11}~\Msun$, centered around the median ($\pm$25\% around median). 
Although $\sigmadla$ increases with increasing $M_h$, the number density of massive halos decreases at the same time.  
This effect pushes the median DLA halo mass to be in the medium mass range. 
\citet{Nagamine07} also presented the median DLA halo mass for various wind models, and our results are similar to their P3 or Q3 model. 
The cyan shade indicates the mass range of DLAs host galaxies suggested by clustering analysis \citep{Cooke06}.
\citet{Cooke06} showed that mass of galaxies hosting DLAs were $10^{9.7} < M < 10^{12}~\Msun$
from DLA-LBG cross-correlation.
Our simulation results agree well with Cooke's result, and the simulation work \citep{Lee11}.  
In addition, we derive the bias parameter from the median mass.
Here we use an ellipsoidal collapse model \citep{Sheth01, Mo02}, 
which can reproduce results of $N$-body simulations well.
We derive the rms fluctuation of mass density at a mass scale M, $\sigma(M)$ by using the median mass of DLAs, and
assign it in the equation (8) of \citet{Sheth01}.
As a result, we obtain the bias of  $b=1.69$ (Table~\ref{table:median}).
This bias is also in the observed range of $1.3 < b_{\rm DLA} < 4$ by \citet{Cooke06}.

The distribution is relatively flat as a function of SFR. 
The typical LBGs have SFR $\approx 1-20~ \Msunyr$, and the fraction of $\sigmadla$ 
contributed by such LBGs is $\sim 30$ per cent based on our simulation. 
The median SFR in the top right panel of Figure~\ref{fig:crossdistdla} is $0.3\,\Msunyr$.
Using the Equation~(1)  in \citet{kennicutt98} for the Salpeter IMF,  
this median SFR corresponds to $\sim 0.02\,L_{\star}$ of LBGs at $z=3$, which is much fainter 
than the observational flux limit ($0.1\,L_{\star}$) in the recent LBG survey of \citet{Reddy09}.  
Therefore we suggest that more than half of total $\sigmadla$ are hosted by the fainter LBGs than the current observational limit,
which is consistent with the observation \citep[e.g.,][]{Fynbo99}.

The green shaded region is an approximate range of SFR for DLAs derived by the $\rm C_{[II]}$ cooling technique
\citep{Wolfe03}.
More than eighty per cent of DLA cross section in our simulations overlaps with the green shaded range. 
There are some observed DLA galaxies with higher SFR of a few tens $\Msunyr$, which are related to LBGs \citep[e.g.,][]{Moller02}.
The massive haloes with $\Mh \gtrsim 10^{11.5}~\Msun$ in our simulation would correspond to those DLA host galaxies with relatively high SFR.

The bottom two panels of Figure~\ref{fig:crossdistdla} show the PDF as functions of stellar mass and gas metallicity. 
The median stellar mass is $\sim 2.4 \times 10^{8} \Msun$, and the median metallicity is $\langle Z/ \Zsun \rangle = 0.1$.
Fifty per cent of total $\sigmadla$ are hosted by galaxies with $3.9 \times 10^{7} < M_\star / \Msun < 1.8 \times 10^{9}$ 
and $0.06 <  Z/ \Zsun < -0.2$, both centered around the median values (i.e. $\pm$25\% around median).    
The distribution in terms of metallicity is strongly peaked around $\langle Z/ \Zsun \rangle = 0.1$, and has an extended tail to lower metallicities. 
Our mean/median metallicity is not so far from the peak of observational data,  $Z/ \Zsun \approx  0.1$ \citep{Prochaska07a}.
However, our PDF is somewhat shifted to higher metallicity than the observed DLA's at $2 \le z \le 4$ \citep{Prochaska07a}
and $1.5 \le z \le 5$ \citep{Rafelski12}. 
Our simulation can successfully reproduce $\fn$, but it seems to overpredict DLA metallicity distribution.
The metallicity--$\NHI$ relation is not sensitive to the wind models \citep[e.g., Figure 9 in ][]{Nagamine04f}.
Hence, our current wind model does not seem to cause the higher metallicity.
The higher metallicity may be caused by the star formation model.
If our star formation model is too efficient at high redshift, the gas is quickly enriched by metals, and it leads to higher metallicity of DLAs.
In addition,  the resolution of numerical simulations may change the metallicity distribution of DLAs.
In general, higher resolution simulations can resolve smaller dwarf galaxies, and many gas clumps in them.
If the gas in the dwarf galaxies which have little metals can contribute to the DLA abundance,
the typical metallicity of DLAs can be reduced.
These topics will be addressed in future works.

Another possibility is the selection bias of DLAs in observation due to the dust reddening \citep[e.g.,][]{Ellison01, Fynbo10, Fynbo11, Khare12}.
Recently \citet{Fynbo10, Fynbo11} found some metal-rich DLAs
using the detection technique based on Si\,{\sc{ii}} and Fe\,{\sc{ii}} absorption lines. 
These objects were not detected in the automatic DLA searches in the SDSS using Ly$\alpha$ \citep[e.g.,][]{Prochaska05}. 
They suggested that dust in such metal-rich DLAs could obscure the QSOs, 
and cause the non-detection of DLAs using Ly$\alpha$.  This would lead to a distribution of DLAs biased to low metallicities. 
Hence, the difference between our simulations and observation may come from this selection bias.
\citet{Khare12} concluded that $\le 10$ per cent of the DLAs in SDSS DR7 cause significant reddening. It is unclear if this reddened population can explain the discrepancy we see in the PDF, but it will bring the two distributions closer. 

The dashed histograms are the PDFs in the optically thin run.
Although $\fn$ significantly changes with the UVB model,
the typical physical quantities do not change very much. 
As we see in Figure~\ref{fig:sigmauv},
$\beta$ in the fitting function (Equation 5) changes with the UVB model,
while the power-law slope $\alpha$ does not change largely.
In our simulations, the physical quantities are roughly proportional to halo mass.
Therefore, the typical physical quantities do not depend on the UVB model very much. 

\begin{figure}
\begin{center}
\includegraphics[scale=0.43]{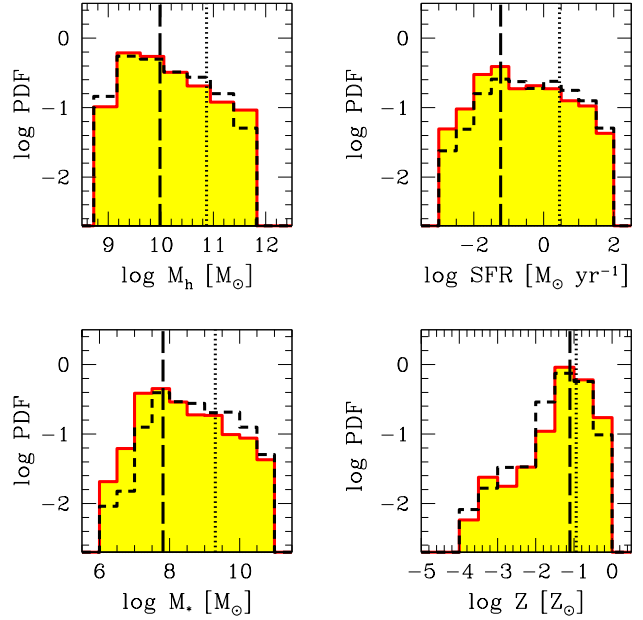}
\caption{
Same as Fig.~\ref{fig:crossdistdla}, but for LLS cross section. 
}
\label{fig:crossdistlls}
\end{center}
\end{figure}

Figure~\ref{fig:crossdistlls} shows the PDF of $\sigmalls$. 
The shape of PDF is basically similar to that of $\sigmadla$. 
However, in the case of LLSs, the contribution from lower mass (and hence lower SFR and lower $\Mstar$) haloes becomes greater than for DLAs. 
The median values are:  $M_h = 9.6 \times 10^{9}\,\Msun$, $\SFR = 0.06\,\Msunyr$, $\Mstar = 6.5 \times 10^{7} \,\Msun$, and $Z/\Zsun = 0.08$.
About 80 per cent of total $\sigmalls$ come from haloes with ${\rm SFR} \lesssim 1~\Msunyr$.
Hence the majority of LLSs are hosted by the low-mass halos with LBGs that are fainter than the current observational limit. 
Fifty per cent of total $\sigmalls$ are hosted by haloes with $2.5 \times 10^{7} < M_\star / \Msun < 5.3 \times 10^{8}$ and $0.05 < Z/ \Zsun <  0.15$. 
We summarize the various median values in Table~\ref{table:median}.

\begin{table}
\begin{center}
\begin{tabular}{ccccccc}
\hline
  &  $\langle \Mh \rangle$ & $\langle {\rm bias} \rangle$ &$\langle {\rm SFR}  \rangle$ & $\langle M_{*}  \rangle$ & $\langle Z / \Zsun \rangle$  \\
   &  $(10^{10}~\Msun)$ && $(\Msunyr)$ & $(10^{8}~\Msun)$ &   \\
\hline
DLA     & 2.39   &1.69& 0.33 & 2.43 & 0.10  \\
LLS   & 0.96 & 1.50&0.06  & 0.65 & 0.08  \\
\hline
\end{tabular}
\caption{
Median values of various physical quantities for the PDF of DLA/LLS cross section. 
The bias is derived from the median halo mass by using an ellipsoidal collapse model \citep{Sheth01}. 
}
\label{table:median}
\end{center}
\end{table}


\section{SUMMARY}
\label{sec:summary}

We examined the physical properties of DLAs and LLSs at $z=3$ using cosmological SPH simulations.
In particular, we studied the impact of radiative transfer of UVB and local stellar radiation on DLAs and LLSs. 
Our major findings are as follows: 

\begin{itemize}
\item The OTUV simulation with a self-shielding density threshold of $\nuv = 6\times 10^{-3}$\,cm$^{-3}$ for the UVB can reproduce the observed $\fn$ very well (Figure~\ref{fig:cdfall}), compared to the run with optically-thin approximation (the Optically-Thin run).  We have presented this result earlier \citep{Nagamine10}, and in this paper we further validated the adopted value of $\nuv$ by the direct RT calculations.  
Similar results have been obtained various authors to a varying degree 
\citep{Pontzen08, Tescari09, Fumagalli11, McQuinn11}.

As we showed in Figs.~\ref{fig:gammauvb}, \& \ref{fig:uvbcom},  the above value of $\nuv$ captures the rapid change in the ionization fraction in halo gas very well.  In the Optically-Thin run, the UVB sinks too deeply into the halo gas and ionizes gas too much, resulting in the underestimate of $\fn$.  
Given that it is still difficult to perform RT calculations simultaneously with hydrodynamics in current cosmological simulations, 
our OTUV model provides a useful prescription for any cosmological simulations that cannot resolve the self-shielding by molecular gas on sub-kpc scales. 

\item We find that the local stellar radiation does not strongly affect the distribution of DLAs/LLSs, and $\fn$ does not change very much by the stellar RT (Figure~\ref{fig:cdfall}).  This is because clumpy high-density clouds near young star clusters effectively absorb most of the ionizing photons from young stars \citep{Yajima11}. 
The effect of stellar RT is stronger on the lower mass halos, and the reduction in $\sigmadla$ and $\sigmalls$ were shown in Figs.~\ref{fig:dlacom}. 

\item We compared the DLA/LLS cross sections in simulations with different UVB models (Figure~\ref{fig:sigmauv}).  In our simulations, the No-UVB run gives the highest DLA/LLS cross sections as expected, and it overpredicts the rate-of-incidence and the column density distribution.  For the DLAs, the OTUV run gives intermediate $\sigmadla$ values, and the HM0.5 \& Optically-Thin run give the lowest $\sigmadla$.  The similarity of the latter two runs tells us that the gas becomes highly ionized under the optically thin run, no matter what the exact UVB amplitude is.  
For the LLSs, the latter three runs (OTUV, Opticall-Thin, and HM0.5) all give similar $\sigmalls$, and the self-shielding model does not matter for LLSs, as they are in the outskirts of galaxies. 

\item We compared the results of DLA cross section from various authors and simulations (Figure~\ref{fig:sigma_com_model}).   There are some general agreement in the sense that most results are bracketed by our Opticall-Thin run and the No-UVB run, and that both $\sigmadla$ and $\sigmalls$ increases with increasing halo mass.   However, there is still significant scatter between different authors.   The origin of the differences could be a combination of many things, such as different models of star formation, galactic wind, UVB and self-shielding treatment.  
For example, the stronger wind model can sweep high-density gas clouds out of star-forming regions,
and decrease DLA cross section.
The self-shielding model of UVB can increase the DLA cross section compared to the optically-thin method,
and rate of increase can be different according to the threshold density for self-shielding.
We have chosen the appropriate value of $\nuv$ based on the good match to the observed $\fn$ and through the validation by direct radiative transfer calculation. 

\item The cumulative number of DLAs per unit redshift strongly depends on the UVB models (Figure~\ref{fig:dndz}). 
The OTUV run agrees well with the observed data.
On the other hand, the No-UVB run overproduces DLAs.
The Optically-Thin and HM0.5 runs underpredict the observation. 

\item The visual appearance of DLAs and LLSs is quite different in the OTUV and Optically-Thin runs (Figure~\ref{fig:column} \& \ref{fig:mtl}). 
In the OTUV run, the DLAs have an extended distribution around star-forming regions on typically $\sim 10-30$\,kpc scales, 
and LLSs are surrounding DLAs on $\sim 30-60$\,kpc scales. 
As shown in the middle row of Figure~\ref{fig:mtl}, the extent of DLAs are typically smaller than the virial radius of the halo, and the LLSs are about the same size as the virial radius.  Given the good agreement between the OTUV run and the observations of $\fn$, we consider that these are the most realistic distribution of DLAs and LLSs in our simulation set. 

On the other hand, in the Optically-Thin run, the UVB penetrates deeply, ionizes the hydrogen gas effectively, and reduces the column density significantly. 
As a result, most of the region around star-forming galaxies turn into LLSs, and the DLAs shrink and become compact.  This striking effect of optically-thin approximation is highlighted by the comparison of columns ($c$) and ($d$) in Figure~\ref{fig:column}, suggesting the importance of self-shielding effect.  The present work supersedes the previous work of \citet{Nagamine04f}, which  reported clumpy distribution of DLA gas in haloes at $z=3$.  With the proper treatment of self-shielding as in the OTUV run, the DLAs have more extended distribution around star-forming galaxies on $10-30$\,kpc scales. 

We quantified the above features in the PDFs of impact parameters between DLAs/LLSs and the nearest galaxies. As expected, we find a much wider distribution for the OTUV run than for the Optically-Thin run.

\item We examined the PDF of DLA/LLS cross sections as functions of halo mass, SFR, stellar mass, and metallicity (Figures~\ref{fig:crossdistdla} and \ref{fig:crossdistlls}). 
The median values of distributions are summarized in Table~\ref{table:median}.  
About 30 per cent of DLAs are hosted by galaxies having $SFR = 1 - 20~\Msunyr$, which is the typical SFR range for LBGs.
More than half of DLAs ($\sigmadla$, to be more precise) are hosted by the LBGs that are fainter than the current observational limit.
About 80 per cent of total $\sigmalls$ are hosted by haloes with ${\rm SFR} \lesssim 1~\Msunyr$, hence 
most LLSs are associated with low-mass halos with faint LBGs below the current detection limit. 

\end{itemize}

%
%
\section*{Acknowledgments}
We are grateful to M. Umemura and J. Miralda-Escud\'{e} for valuable discussion and comments, 
and to Volker Springel for providing us with the original version of 
GADGET-3, on which Choi \& Nagamine simulations are based.
We thank the anonymous referee for useful comments.
This work is supported in part by the NSF grant AST-0807491, 
National Aeronautics and Space Administration under Grant/Cooperative 
Agreement No. NNX08AE57A issued by the Nevada NASA EPSCoR program, and 
the President's Infrastructure Award from UNLV. 
Support for Program number HST-AR-12143.01-A was provided by NASA through a grant from the Space Telescope Science Institute, which is operated by the Association of Universities for Research in Astronomy, Incorporated, under NASA contract NAS5-26555.
This research is also supported by the NSF through the TeraGrid resources 
provided by the Texas Advanced Computing Center (TACC).
Numerical simulations and analyses have been performed on the UNLV 
Cosmology Cluster, the {\it FIRST} simulator  and {\it T2K-Tsukuba} at 
Center for Computational Sciences, University of Tsukuba. 
KN acknowledges the hospitality and the partial support from the Kavli Institute for Physics and Mathematics of the Universe (IPMU), University of Tokyo, the Aspen Center for Physics, and the National Science Foundation Grant No. 1066293.

%
%




\label{lastpage}

\end{document}